\documentclass[prd,twocolumn,aps,prc,nofootinbib,showpacs,showkeys]{revtex4}
\usepackage{epsfig}
\usepackage{graphicx}
\usepackage{amsfonts}
\usepackage{amsmath}
\usepackage{xcolor}

\definecolor{red}{rgb}{1,0,0}
\definecolor{green}{rgb}{0,1,0}
\definecolor{blue}{rgb}{0,0,1}

\usepackage{float}

\newcommand{\pipi}{$\pi\pi$ }
\newcommand{\kk}{$K \bar K$ }
\newcommand{\KK}{$K \bar K$ }

\newcommand{\fsig}{$f_0(500)$ }

\begin{document}

\title{Dispersive analysis of the \mbox{\boldmath $S$}-, \mbox{\boldmath $P$}-, 
\mbox{\boldmath $D$}-, and \mbox{\boldmath $F$}-wave \mbox{\boldmath $\pi\pi$} amplitudes}

\author{P. Byd\v{z}ovsk\'y$^a$, R. Kami\'nski$^b$, V. Nazari$^b$}
\affiliation{$^a$Nuclear Physics Institute, Czech Academy of Sciences, \v{R}e\v{z}, Czech Republic\\
$^b$Institute of Nuclear Physics, Polish Academy of Sciences, Krak\'ow, Poland}

\date{\today }

\begin{abstract}
A reanalysis of \pipi amplitudes for all important partial-waves below 
about 2~GeV is presented. A set of once subtracted dispersion relations 
with imposed crossing symmetry condition is used to modify unitary multi-channel 
amplitudes in the $S$, $P$, $D$, and $F$ waves. 
So far, these specific amplitudes 
constructed in our works and many other analyzes have been fitted only 
to experimental data and therefore do not fulfill the crossing symmetry condition. 
In the present analysis, the self consistent, i.e. unitary and fulfilling 
the crossing symmetry, amplitudes for the $S$, $P$, $D$, and $F$ waves 
are formed.
The proposed very effective and simple method of modification of the \pipi 
amplitudes does not change their previous-original mathematical structure and 
the method can be easily applied in various other analyzes.
\end{abstract}



\maketitle

\section{Introduction}

The enthusiasm in the analysis of \pipi interaction amplitudes has been increased 
significantly quite recently. 
Especially important were numerous works on dispersive analyzes of experimental 
data made by Bern~\cite{A4} and Madrid-Krak\'ow~\cite{GKPY} group.
The significant progress was made when the analyzes began to effectively use theoretical 
constrains i.e. crossing symmetry condition imposed on the amplitudes found in experimental 
analyzes. This was particularly important because of big differences between results obtained 
by various experimental groups and even between data sets found in the same experimental 
analysis~\cite{PY05,RKProceedings}.

Those dispersive analyzes had immediately large impact on the spectroscopy of light scalar mesons (i.e. $f_0(500)$ and $f_0(980)$) which is evident comparing the tables of the Particle Data Group from the years 2010 and 2012 \cite{PDG2010,PDG2012}. 
The analyzes provided also a set of all important amplitudes ($S$, $P$, $D$ and $F$) well describing experimental data up to 1420~MeV and 2000~MeV in case of works done by the Madrid-Krak\'ow and Bern group, respectively.
In both those analyzes the $S$ and $P$ wave amplitudes were fitted also to dispersion relations with imposed crossing symmetry constrain below 1100~MeV. 
The Bern group was using Roy equations \cite{Roy1971} which need two subtractions and 
found analytical solution below 800~MeV.
The Madrid-Krak\'ow group also used Roy equations and additionally Roy like ones, so called GKPY equations with only one subtraction what gave more precise output amplitudes.
The amplitudes of higher partial-waves, $D$ and $F$, were also fitted to such 
dispersion relations but only indirectly by means of the Roy or GKPY equations, 
i.e. they were present in the kernel part of the equations written for the $S$ 
and $P$ waves.

An example of practical application of the GKPY equations is our last reanalysis 
of the $S$- and $P$-wave multi-channel amplitudes~\cite{BKN2014} constructed in 
a previous analysis by fitting the experimental data only~\cite{Yura2010}.
One of the most spectacular effects given by the GKPY equations was a shift 
of the \fsig pole by several hundred MeV towards the position indicated by 
the analyzes of the Bern and Madrid-Krak\'ow group.
It is important to note that this reanalysis was done keeping the original 
mathematical structure of the amplitudes proposed in the work~\cite{Yura2010}.

The aim of this work is to perform similar but more extensive reanalysis of 
the multi-channel \pipi amplitudes including the $S$, $P$, $D$, and $F$ partial 
waves important in the low energy region (below 2~GeV). 
The initial amplitudes, in the following denoted as 
\textquotedblleft original\textquotedblright, were fitted only to 
the experimental data and are taken from our previous analysis 
($S0$~\cite{BKN2014} and $P1$~\cite{Yura2010}, hereafter we will use 
notation $\ell I$ if needed ($\ell$-meson-meson partial wave and $I$-isospin)) 
or are updated in this 
work utilizing a form from~\cite{YuraDF} ($D0$ and $F1$). 
To perform this analysis we use the GKPY equations for the $D$ and $F$  
waves constructed and presented in \cite{osdrdf}, which have not been used so 
far in the analysis of amplitudes. 
The final amplitudes are constrained by both the experimental data 
and the GKPY dispersion relations (i.e. by crossing symmetry). As the reanalysis  
is not too much effective for higher partial-waves, particularly for $F1$, 
parameters of this amplitude are expected to be only weakly changed. 
In the analysis, only some parameters of the amplitudes are changed which 
do not alter the mathematical structure of the original amplitudes, similarly 
as in our previous analysis of the $S$ and $P$ waves \cite{BKN2014}.

The reanalyzed partial-wave multi-channel amplitudes, constrained also by the 
crossing symmetry, can be utilized in accounting for the final-state interactions 
in the decays of heavy mesons and in photoproduction processes, for example, 
in the CLAS12 and GlueX experiments at JLab. 
The amplitudes can also be used in constructing 
a full \pipi amplitude giving the cross section at low energies.   

The paper is organized as follows: in Section II we briefly remind the method 
and results of our previous analysis of the $S0$ and $P1$ partial-wave 
amplitudes  done in~\cite{BKN2014}. In Section III we present new $D0$ and 
$F1$ amplitudes fitted to experimental data. 
In Section IV we detail the method and give results of the dispersive 
analysis for all considered partial-waves. 
Section V is devoted to discussion of obtained results. 
Here we also show results for the low-energy total and differential 
cross sections in the $\pi^+\pi^-$ scattering.
Summary of results is in Section VI. 

\section{Dispersive analysis of the \mbox{\boldmath $S$}- and \mbox{\boldmath $P$}- wave amplitudes}
\label{SPanalysis}

In our previous work~\cite{BKN2014} we reanalyzed the $S$- and $P$-wave amplitudes 
constructed in~\cite{Yura2010} refitting some of their parameters simultaneously 
to experimental data and to GKPY dispersion relations. These multi-channel amplitudes 
are unitary and analytic on the Riemann surface with free parameters that are 
just positions of poles on different Riemann sheets and few background parameters. 
Such simple and not biased mathematical structure makes interpretation of obtained 
results very easy and unambiguous.
In analysis \cite{Yura2010} these amplitudes were, however, fitted only to very 
dispersed experimental data 
from various experiments what resulted in pole positions sometimes 
very different from those obtained in other analyzes and from those in Particle 
Data Group Tables.

In the analysis \cite{BKN2014} the original mathematical structure of 
the resonant and background parts of amplitudes from \cite{Yura2010} was not 
changed. The only novelty was a new parameterization of the near threshold 
amplitudes, which was necessary due to the total lack of description of the 
phase shifts in this region~\cite{Yura2010}.
Polynomials were, therefore, added to both $S$ and $P$ amplitudes and 
the phase shifts (values and the first derivatives) were smoothly matched 
below the \kk threshold, at about 400 and 600~MeV for the $S$ and $P$ waves, 
respectively.
In Ref.~\cite{BKN2014} we also constructed a ``new'' $S$-wave isoscalar amplitude 
hereafter called New $S$-wave fitting its parameters only to the data, what 
improved behavior of the amplitude. In the subsequent analysis 
both ``Old''~\cite{Yura2010} and New $S$-wave isoscalar amplitudes were used. 

In analysis \cite{BKN2014} the strategy of our work was following: 
the Old (New) $S$-wave isoscalar and 
the old isovector $P$-wave~\cite{Yura2010} amplitudes, both supplemented with  
the near threshold polynomials, were used as the input in the GKPY equations 
that have a general form
\begin{eqnarray}\nonumber 
{\mbox{Re } t_{\ell}^{I^{(OUT)}}(s)} & = & \sum\limits_{I'=0 }^2 C^{II'}{ t^{I^\prime}_0(4m_{\pi}^2)} \\  
&\hspace{-1.7cm} + &  
\hspace{-1cm} \displaystyle \sum\limits_{I'=0}^2
        \displaystyle \sum\limits_{\ell'=0}^3
     \hspace{0.2cm}-\hspace{-0.48cm}
        \displaystyle \int \limits_{4m_{\pi}^2}^{\infty}\ \hspace{-0.2cm}ds'
     K_{\ell \ell^\prime}^{I I^\prime}(s,s')\, {\mbox{Im }t_{\ell'}^{I^{\prime \,(IN)}}(s')}\,,\ \ \ 
\label{EqGKPY}
\end{eqnarray}
where $t_{\ell^{\prime}}^{I^{\prime \,(IN)}}(s')$ and $t_{\ell}^{I^{(OUT)}}(s)$ 
are the input and output amplitudes, respectively, in a given partial-wave 
$\ell, \ell^\prime$ with isospin $I, I^\prime$. The $C^{II'}$ is the crossing 
matrix constant and $K_{\ell \ell^\prime}^{I I^\prime}(s,s')$ are kernels 
constructed for partial-wave projected amplitudes with the imposed $s 
\leftrightarrow t$ crossing symmetry condition. 
The kernels for the $S$ and $P$ waves were presented in \cite{GKPY}.

As it is seen in Eq. (\ref{EqGKPY}), one has to use   
the imaginary parts of all important partial-wave amplitudes as an input. 
We took, therefore, these important amplitudes, $S2$, $D0$, $D2$ and $F1$, directly from 
Ref.~\cite{GKPY} and kept them fixed during the analysis. 
Finally we fitted some parameters of the $S0$ and $P1$ amplitudes simultaneously 
to the experimental data and to the dispersion relations (\ref{EqGKPY}).  
As a result, the \fsig pole moved by several hundred MeV to a new position 
close to that found in dispersive analyzes \cite{ccl,gar-mar}. 

The minimized full $\chi^2$ function was composed 
of two data terms $\chi^2_{Data}(k)$, related with data in the $S$ 
and $P$ waves, and of three terms $\chi^2_{DR}(k)$ for the output amplitudes 
from the dispersion relations
\begin{equation}
\label{EqChi2Total}
\chi^2 = \sum_{k=1}^{2}\chi^2_{Data}(k) + \sum_{k=1}^{3}\chi^2_{DR}(k) ,
\end{equation}
where $k = 1, 2, 3$ itemizes, respectively $\ell I$ partial-waves: the $S0$, $P1$ and $S2$.
Corresponding $\chi^2_{Data}(k)$ and  $\chi^2_{DR}(k)$ were expressed by
\begin{equation}
\label{EqChi2Data}
\chi^2_{Data}(k) = 
\sum_{i=1}^{N_{\delta}^k} \frac{(\delta_i^{exp}-\delta_i^{th})^2}
{(\Delta \delta_i^{exp})^2} + 
\sum_{i=1}^{N_{\eta}^k} \frac{(\eta_i^{exp}-\eta_i^{th})^2}
{(\Delta \eta_i^{exp})^2}
\end{equation}
and
\begin{equation}
\label{EqChi2DR}
\chi^2_{DR}(k) = \sum_{i=1}^{N_{DR}} \frac{\left[{\rm Re}\: t_{\ell}^{I^{(OUT)}}(s_i)-
{\rm Re}\:t_{\ell}^{I^{(IN)}}(s_i)\right]^2}{\left[\Delta {\rm Re}\: t_{\ell}^{I^{(OUT)}}(s_i)\right]^2},
\end{equation}
where $\delta_i^{exp}$ (or $\eta_i^{exp}$) and $\delta_i^{th}$ (or $\eta_i^{th}$) are experimental and our-theoretical phase shifts (or inelasticities) and $N_{\delta}^k$ (or $N_{\eta}^k$) are numbers of the data points for phase shifts (or inelasticities) of the $S0$ and $P1$ partial-waves in considered coupled channels. 
Symbol $N_{DR}$ is a number of energy points between the \pipi threshold and 1100~MeV, at which we calculated $\chi^2_{DR}(k)$ (for all three waves $N_{DR}=26$ was chosen) and $\Delta {\rm Re}\: t_{\ell}^{I^{(OUT)}}(s_i)$ are fixed to 0.01 in order to make the $\chi^2_{DR}(k)$ comparable with the $\chi^2_{Data}(k)$.

Let us here notice that terms $\chi^2_{DR}(k)$ are in fact not $\chi^2$ functions but rather squared weighted differences between the input and output amplitudes. For simplicity we keep, however, the name $\chi^2$ for these terms.

\section{New analysis of the \mbox{\boldmath $D$}- and 
\mbox{\boldmath $F$}-wave experimental data}

The experimental data for the $\pi\pi$ scattering in the $D0$ and $F1$ waves 
were analyzed in Refs.~\cite{Yura2010,YuraDF} to study the $f_2$ and $\rho_3$ 
mesons. 
In presented here analysis the S-matrix formalism for N coupled 
channels was utilized similarly as in our previous dispersive analysis of $S$ 
and $P$ waves~\cite{BKN2014}. Due to the large number of opened channels in 
the $D$ and $F$ waves the uniformizing variable (see Eq. (1) in \cite{BKN2014}) 
could not be used and therefore, the Jost matrix determinant was constructed 
using the multi-channel Breit-Wigner forms. 

In the present analysis we have used the same formalism for the $D0$ and $F1$ 
waves as in Refs.~\cite{Yura2010,YuraDF} and updated the list of contributing 
resonance states for the $D0$ wave according to the latest issue of 
PDG~\cite{PDG2014}. The corresponding free parameters were fitted to experimental 
data. In the case of $F1$ wave we have found that enough is only one resonance 
state $\rho_3{(1690)}$ and have constructed a reasonable description also in the 
threshold region. These updated $D0$ and $F1$ amplitudes were then used in the 
analysis with the GKPY equations taken from \cite{osdrdf} (see \ref{SecresDF}). 
In the following subsections we give more details on the formalism and 
construction of the New $D0$ and New $F1$ amplitudes.

\subsection{Formalism}

The matrix elements $S_{ij}$ of the $N$-channel $S$ matrix ($i,j=1,2,...N$) are 
expressed via the Jost matrix determinant, $d(k_1,k_2,...k_N)$ ($k_i$ are the 
channel momenta), using the Le Couteur--Newton relations~\cite{Yura2010,YuraDF}. 
These expressions together with the formulas of analytical continuation of the 
matrix elements to the unphysical sheets naturally generate the resonance 
poles and zeros on the Riemann surface.  
The Jost determinant is considered in a separable form $d = d_{bgr}\,d_{res}$.  
The resonance part is described by the multi-channel Breit-Wigner form
\begin{equation}
\label{JostDet_res}
d_{res} = \prod_r\left[ M^2_r - s -i\sum_{j=1}^N\rho^{2J+1}_{rj}R_{rj}f^2_{rj} 
\theta(s-s_j)\right] ,
\end{equation}
where $s$ is the invariant total energy squared, $M_r$ and $J=\ell$ are the resonance 
mass and spin, respectively,  $\rho_{rj}=2k_j/\sqrt{M^2_r-s_j}$ with $s_j$ 
the channel thresholds, $R_{rj}$  are the Blatt-Weisskopf barrier factors, 
and the free parameter $f_{rj}$ is related with a decay width of a resonance $r$ 
into a channel~$j$.
 
The background part $d_{bgr}$, which represents mainly an influence of neglected 
channels and resonances, adds in general an energy dependent phase in each channel. 

\subsection{Fits for the \mbox{\boldmath $D$} wave}\label{fitsDwave}  

In the analysis of the data in the tensor-isoscalar sector we have considered 
explicitly the channels: 1- $\pi\pi$, 2- effective $(2\pi)(2\pi)$, 3- $K\bar{K}$, 
and 4- $\eta\eta$. The resonant part of the Jost determinant, $d(k_1,k_2,k_3,k_4)$, 
is then given by the four-channel Breit-Wigner form (\ref{JostDet_res}) with 
$J=2$ and the barrier factor 
\begin{equation}
\label{BW_factor}
R_{rj}= \frac{9+\frac{3}{4}(\sqrt{M^2_r-s_j}r_{rj})^2+
\frac{1}{16}(\sqrt{M^2_r-s_j}r_{rj})^4}
{9+\frac{3}{4}(\sqrt{s-s_j}r_{rj})^2+\frac{1}{16}(\sqrt{s-s_j}r_{rj})^4} ,
\end{equation}
where the radii $r_{rj}$ have a common value 0.943 fm \cite{YuraDF} which was 
kept constant in our analysis.

In the set of resonance states contributing to the process we have considered 
eleven states presented in the PDG summary table~\cite{PDG2014}: 
$f_2(1270)$, $f_2(1430)$, $f_2(1525)$, $f_2(1640)$, $f_2(1810)$, $f_2(1910)$, 
$f_2(1950)$, $f_2(2010)$, $f_2(2150)$, $f_2(2300)$, $f_2(2340)$. We have not included 
the broad state $f_2(1565)$ which was not listed in the previous issue of PDG 
and which can be mimic by the nearby state $f_2(1525)$. 
The masses of the resonances were taken from the PDG tables but in the curse 
of fitting $f_2(1430)$, $f_2(1525)$, $f_2(2010)$, $f_2(2300)$ and $f_2(2340)$ 
resonance masses were allowed to change slightly within an interval of several 
standard deviations around the central value. The partial widths of the resonances, 
the parameters $f_{rj}$ in (\ref{JostDet_res}), were fitted to the data.

The background part of the Jost determinant was taken from~\cite{Yura2010} 
and has the form 
\begin{equation}
\label{JostDet_bgrD}
d_{bgr} = \exp \left[ -i\sum_{j=1}^4\left(\frac{2k_j}{\sqrt{s}}\right)^{5}
(a_j+ib_j) \right] ,
\end{equation}   
where $a_2=a_3=a_4=0$,  
$$ a_1= \alpha_{11}+\frac{s-s_3}{s}\alpha_{13}\,\theta(s-s_3) +
\frac{s-s_v}{s}\alpha_{10}\,\theta(s-s_v) ,
$$
$$ b_j= \beta_j+\frac{s-s_v}{s}\gamma_j\,\theta(s-s_v),\ \ {\rm for}\ \ j=1,3,4\;,$$ 
and $b_2=0.$
The threshold $s_v=2.274$~GeV$^2$ accounts for effects from the channels 
$\eta\eta', \rho\rho$, and $\omega\omega$ not included explicitly in the analysis.  
The parameters $\alpha_{11},\,\alpha_{13},\,\alpha_{10},\,\beta_j,$ and $\gamma_j$
were fitted to the data.

The experimental data for the $\pi\pi$ scattering are from the energy-independent 
analysis by Hyams et al.~\cite{Hyams} and the data for inelastic scattering 
$\pi\pi \to K\bar{K}, \eta\eta$ from Ref.~\cite{Lindenbaum}. 
To warrant a right behavior of the elastic phase shifts in the threshold region, 
i.e. a consistency with the scattering length and the slope parameter taken 
from \cite{GKPY}, we have included in the data set also the sixteen points 
(pseudo data) generated in the range 282-825~MeV by the phenomenological 
amplitudes~\cite{GKPY} with errors about 10\%. 

\begin{figure}[h!]
\includegraphics[angle=270,scale=0.31]{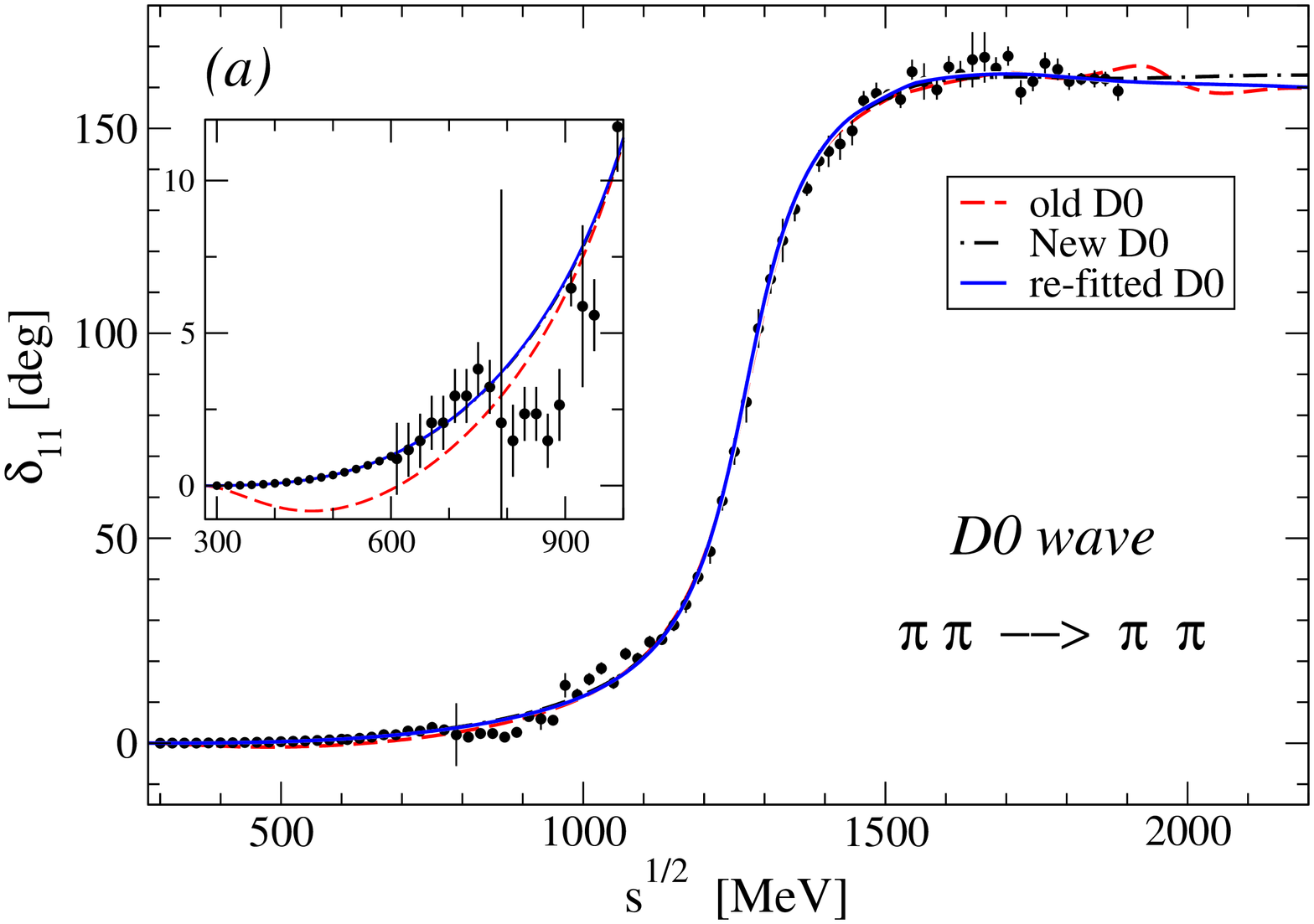}\\
\includegraphics[angle=270,scale=0.31]{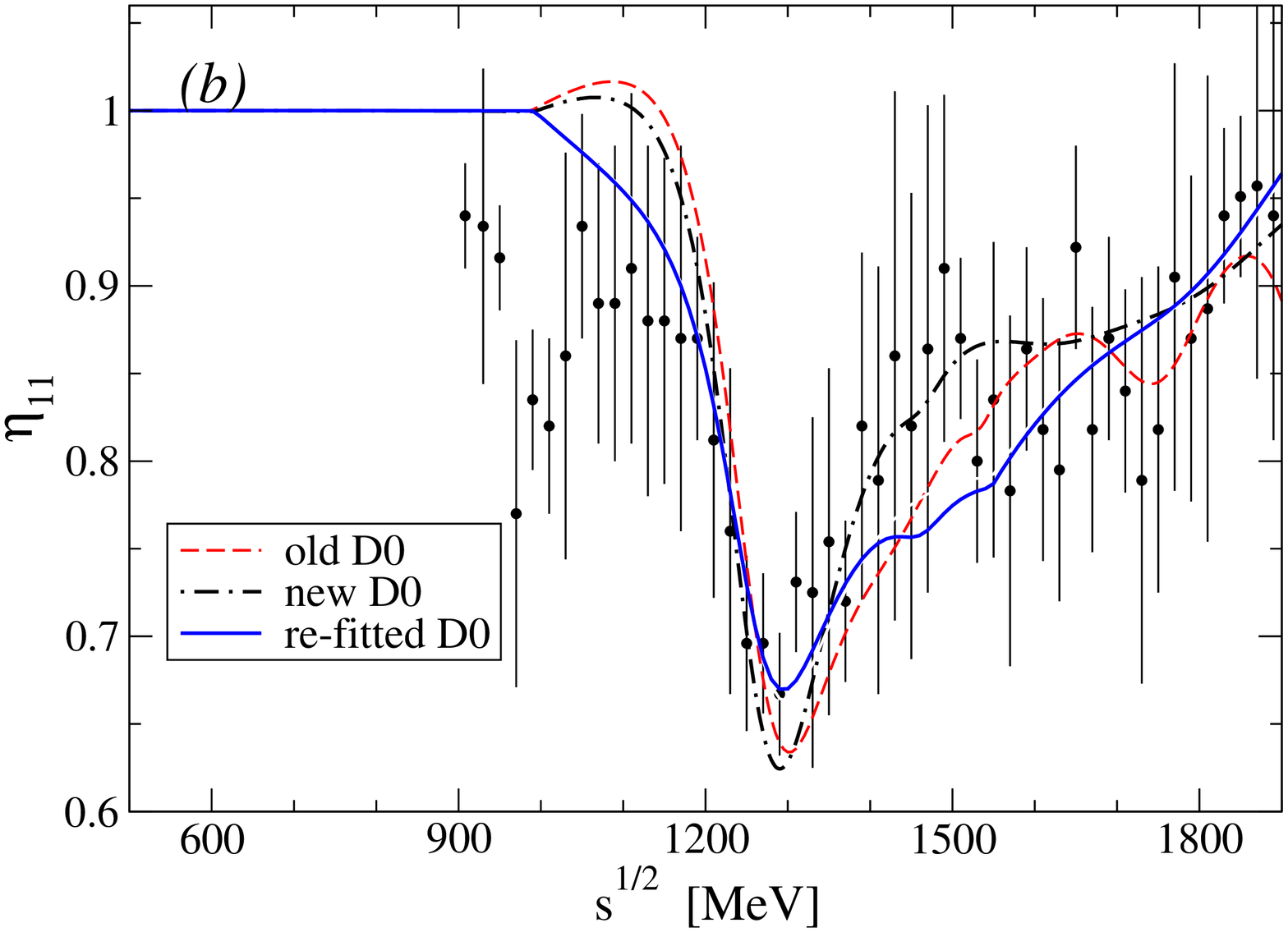}
\caption{Results of the New $D0$ (dashed-dotted line), old $D0$~\cite{Yura2010} 
(dashed line) and re-fitted (after fitting to data and dispersion relations) 
$D0$ (solid line) amplitudes for the phase shift ($a$) and inelasticity ($b$) 
of the $\pi\pi \to \pi\pi$ scattering are compared with experimental data from 
Ref.~\cite{Hyams}  and the pseudo data.}
\label{figD0-1}
\end{figure}

\begin{figure}[h!]
\includegraphics[angle=270,scale=0.31]{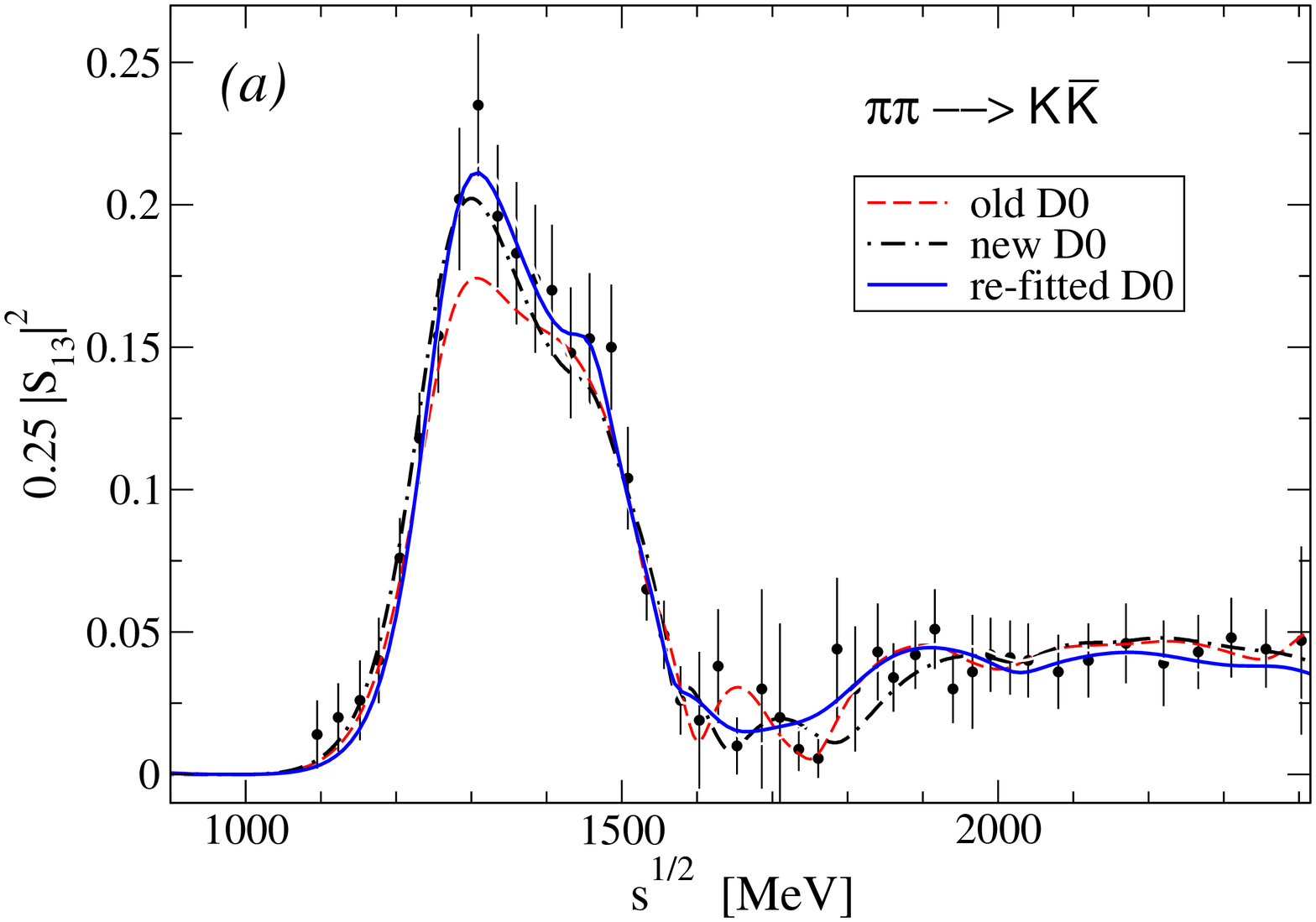}\\
\includegraphics[angle=270,scale=0.31]{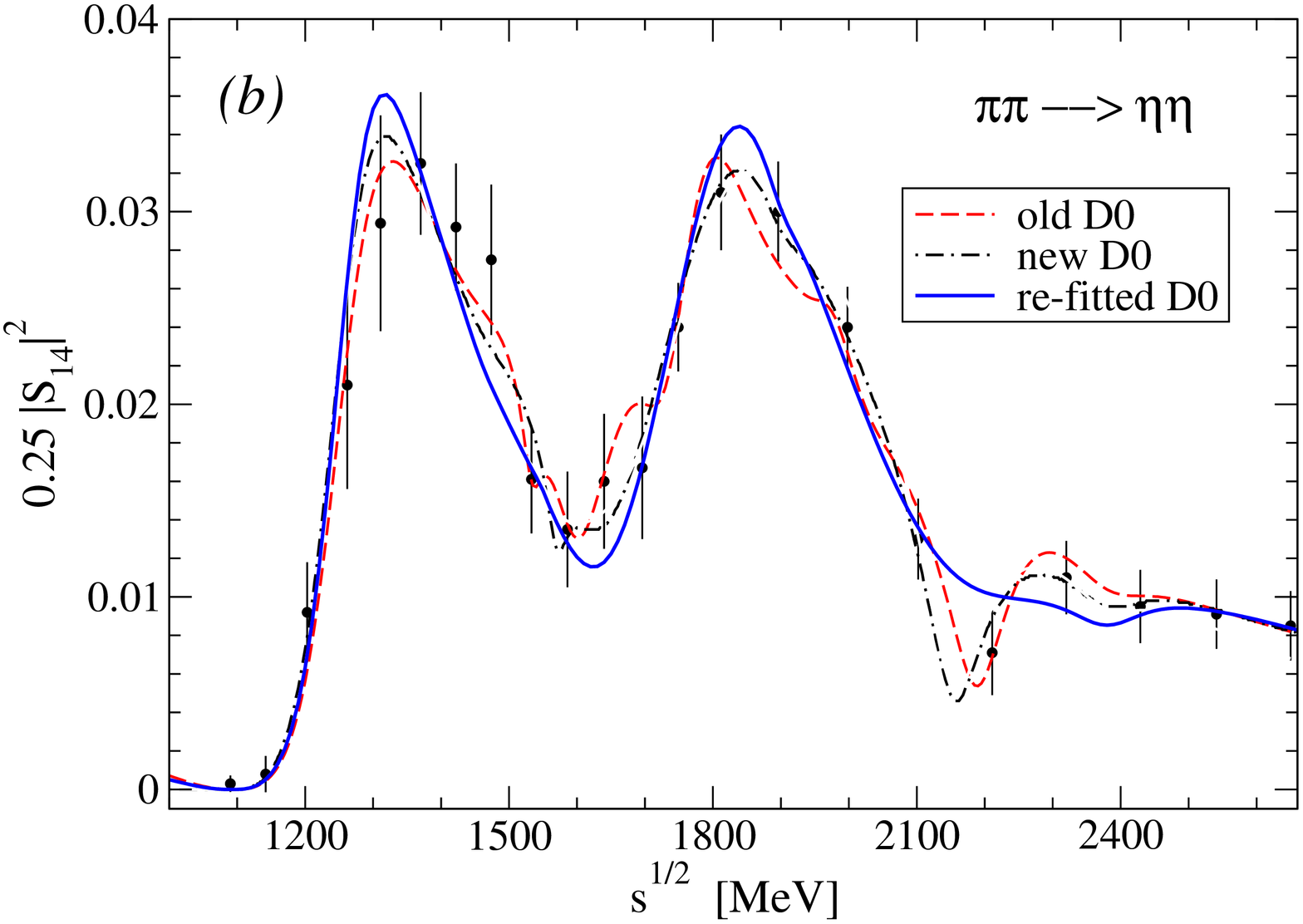}
\caption{The same as in Fig.~\ref{figD0-1} for the squared modulus of the S matrix 
for the $\pi\pi \to K\bar{K}$ ($a$) and $\pi\pi \to \eta\eta$ ($b$).}
\label{figD0-2}
\end{figure}

In successive fitting of the parameters to the data we found a solution 
with $\chi^2/n.d.f. = 242.28/(199 -58) = 1.72$ where the value without 
the pseudo data is $\chi^2/n.d.f. = 239.28/(183 -58) = 1.91$.  
The parameters of resonances of the New $D0$ amplitude are given  
in Table~\ref{newD0} and a comparison with the old $D0$ amplitude~\cite{Yura2010} 
is shown in Figs.~\ref{figD0-1} and \ref{figD0-2}. 
%
%
\begin{table}
\begin{tabular}{ccccccc}  
\hline 
 state & PDG     &  $M_r$  & $f_{r1}$& $f_{r2}$& $f_{r3}$& $f_{r4}$ \\  
\hline 
$f_2(1270)$& 1275.5 $\pm$ 0.8 & 1275.5 & 459.3  & 0.001 & 204.0 &  91.3 \\
$f_2(1430)$& 1430             & 1463.2 &  42.3  & 0.12  & 346.8 & 0.02 \\
$f_2(1525)$& 1525 $\pm$ 5     & 1570.7 &  0.01  & 207.5 & 128.4 &  96.3 \\
$f_2(1640)$& 1639 $\pm$ 6     & 1639.0 & 145.3  & 524.4 & 430.5 & 233.5 \\
$f_2(1810)$& 1815 $\pm$ 12    & 1815.0 & 163.5  & 279.2 & 497.2 & 590.3 \\
$f_2(1910)$& 1903 $\pm$ 9     & 1903.0 &  0.077 &  65.3 & 0.067 & 371.3 \\
$f_2(1950)$& 1944 $\pm$ 12    & 1944.0 &  5.01  &  59.5 & 625.7 &  97.9 \\
$f_2(2010)$& 2011 $\pm$ 62    & 2027.0 &  0.001 & 146.4 & 457.1 &   0.5 \\
$f_2(2150)$& 2157 $\pm$ 12    & 2157.0 &  0.015 & 445.8 & 148.1 & 354.6 \\
$f_2(2300)$& 2297 $\pm$ 28    & 2181.6 &  78.14 &  74.9 & 818.3 & 169.5 \\
$f_2(2340)$& 2345 $\pm$ 40    & 2383.3 &  46.20 &   7.1 & 633.2 & 163.8 \\
  \hline 
\end{tabular}
\caption{Parameters of the Breit-Wigner form (in MeV) for the New $D0$ amplitude. 
The masses of the resonances from PDG~\cite{PDG2014} are also shown in the second 
column.} 
\label{newD0}
\end{table}
The background parameters are: $\alpha_{11}= 0.00096$, $\alpha_{13}=-0.04105$, 
$\alpha_{10}=-0.186$, $\beta_1 = -0.0531$, $\beta_3 = -1.99$, 
$\beta_4 = -1.47$, $\gamma_1 = 0.00128$, $\gamma_3 = 1.99$, and $\gamma_4 = 1.43$. 

This solution is a bit worse (the $\chi^2$) than that in Ref.~\cite{Yura2010} 
($\chi^2$ was $156.62/(168 -69) = 1.58$) but we have achieved the right behavior 
of $\delta_{11}$ for energies below 800~MeV (see the detail in Fig.~\ref{figD0-1}($a$)) 
which allows us to avoid a polynomial-like extension 
of the phase shift as in the case of the $S$ and $P$ amplitudes. 
In Figs.~\ref{figD0-1} and \ref{figD0-2} for inelasticity $\eta_{11}$ and  
the squared modulus of the S matrix in inelastic channels one can see even 
a slight improvement in description of the data. Please notice also that 
the set of resonances and their masses are in a good agreement 
with the PDG tables, see Table~\ref{newD0}.

%
%

%
%
  
\subsection{Fits for the \mbox{\boldmath $F$} wave}\label{fitsFwave}

In the isovector $F$ wave, there are only two resonance states listed in the PDG 
summary table which are relevant for the data description below 2~GeV: 
$\rho_3(1690)$ and $\rho_3(1990)$~\cite{PDG2014}. For the former, the decay widths 
into the $\pi\pi$, $\pi^\pm\pi^+\pi^-\pi^0$, $\omega\pi$, $K\bar{K}$ and $K\bar{K}\pi$ channels 
are well established whereas for the latter the partial widths are not known. 
To learn on importance of these resonances in description of data we performed 
fits in~\cite{DFVN} (Tables V and VI). We showed that if both resonances 
are fitted simultaneously then the mass of $\rho_3(1990)$ turns into a huge number 
showing that one resonance state is enough to achieve a reasonable data 
description.  
We have therefore considered only $\rho_3(1690)$ in the analysis of the phase 
shift and inelasticity parameter in the $\pi\pi$ scattering~\cite{Hyams}. 
This state is also apparently well seen in the pronounced data structure.

In the analysis we have included four channels: 1- $\pi\pi$, 2- effective 
$(2\pi)(2\pi)$, 3- $\omega\pi$, and 4- $K\bar{K}$. The Blatt-Weisskopf barrier 
factor in the Breit-Wigner form (\ref{JostDet_res}) was in the case of $J=3$
\begin{equation}
\label{BW_factor_F}
R_{1j}= \frac{225 +45(\mu_jr_{1j})^2 +6(\mu_jr_{1j})^4 
+(\mu_jr_{1j})^6}
{225 +45(2k_jr_{1j})^2 +6(2k_jr_{1j})^4 +(2k_jr_{1j})^6} ,
\end{equation}
where $\mu_j$ is equal to $\sqrt{M^2_1-s_j}$. The radii $r_{1j}$ possess 
a common value 0.927 fm \cite{YuraDF} that were kept constant in our analysis. 
The resonance mass and the Breit-Wigner parameters $f_{1j}$ were fitted to the data.

In our analysis the dispersion relations directly affect only the energy region 
below 1100~MeV but the higher energy region, where the $\rho_3(1690)$ resonance 
is clearly seen, is influenced indirectly. The $F1$ amplitude is, therefore, 
almost entirely determined by the experimental data. Since the resonance state  
is described by the Breit-Wigner form, which works well only in a vicinity of 
the resonance, we had to take a particular care of behavior of the amplitude 
below about 1000~MeV. We have, therefore, chosen a simple modification of the 
phase shift by means of the background phase in the form of the quadratic polynomial 
of $s$
\begin{eqnarray}
d_{bgr}= \exp\Big[-i\Big(\frac{2k_1}{\sqrt{s}}\Big)^7\Big(a_{\alpha} + 
\frac{4k^2_1}{s_1}a_{\beta} + (\frac{4k^2_1}{s_1})^2a_{\gamma} \Big)\Big] ,
\label{bgr_F1}
\end{eqnarray}
where $k_1$ is equal to $\sqrt{s-s_1}/2$ and the parameters $a_\alpha$, 
$a_\beta$, and $a_\gamma$ were fitted to the data.
Similarly as in the case of the $D0$ wave 
we have included pseudo data points but now 31 for energies 282~MeV $<\sqrt{s}<$ 895~MeV. 
These data were generated by the phenomenological parameterization of 
the amplitudes~\cite{GKPY}, which ensure the right values of the scattering 
length and slope parameter given in~\cite{GKPY}.

In the fitting with the four-channel form we found that only the channels 1 and 2 
are important whereas the parameters $f_{13}$ and $f_{14}$ were almost zero. 
This corresponds quite well with the observed decay rates of the $\rho_3(1690)$ 
resonance: 23.6$\pm$1.3\%, 67$\pm$22\%,  
16$\pm$6\%, 3.8$\pm$1.2\%, and 1.58$\pm$0.26\% into the 
$\pi\pi$, $\pi^\pm\pi^+\pi^-\pi^0$, $\omega\pi$, $K\bar{K}\pi$, and $K\bar{K}$ 
channels, respectively~\cite{PDG2014}.
We did therefore a two-channel fit with $\chi^2/{n.d.f.}= 136.20/(108-6) = 1.34$ 
where the value without the pseudo data is $\chi^2/{n.d.f.}= 80.84/(77-6) = 1.38$ 
showing that the pseudo data are quite consistent with the experimental 
data used in the analysis. 
The resonance parameters are shown in Table~\ref{newF1} and the background parameters 
are $a_\alpha = 0.000008$, $a_\beta = -0.000998$, and $a_\gamma = 0.000016$. 
These parameters are quite small but they play an important role. In the fit 
with only the first term in (\ref{bgr_F1}) the 
$\chi^2/{n.d.f.} = 4.0$ where the main contribution comes from the pseudo data. 
This we consider as a strong evidence of the need for the nonzero additional 
terms in (\ref{bgr_F1}).

\begin{figure}[h!]
\includegraphics[angle=270,scale=0.31]{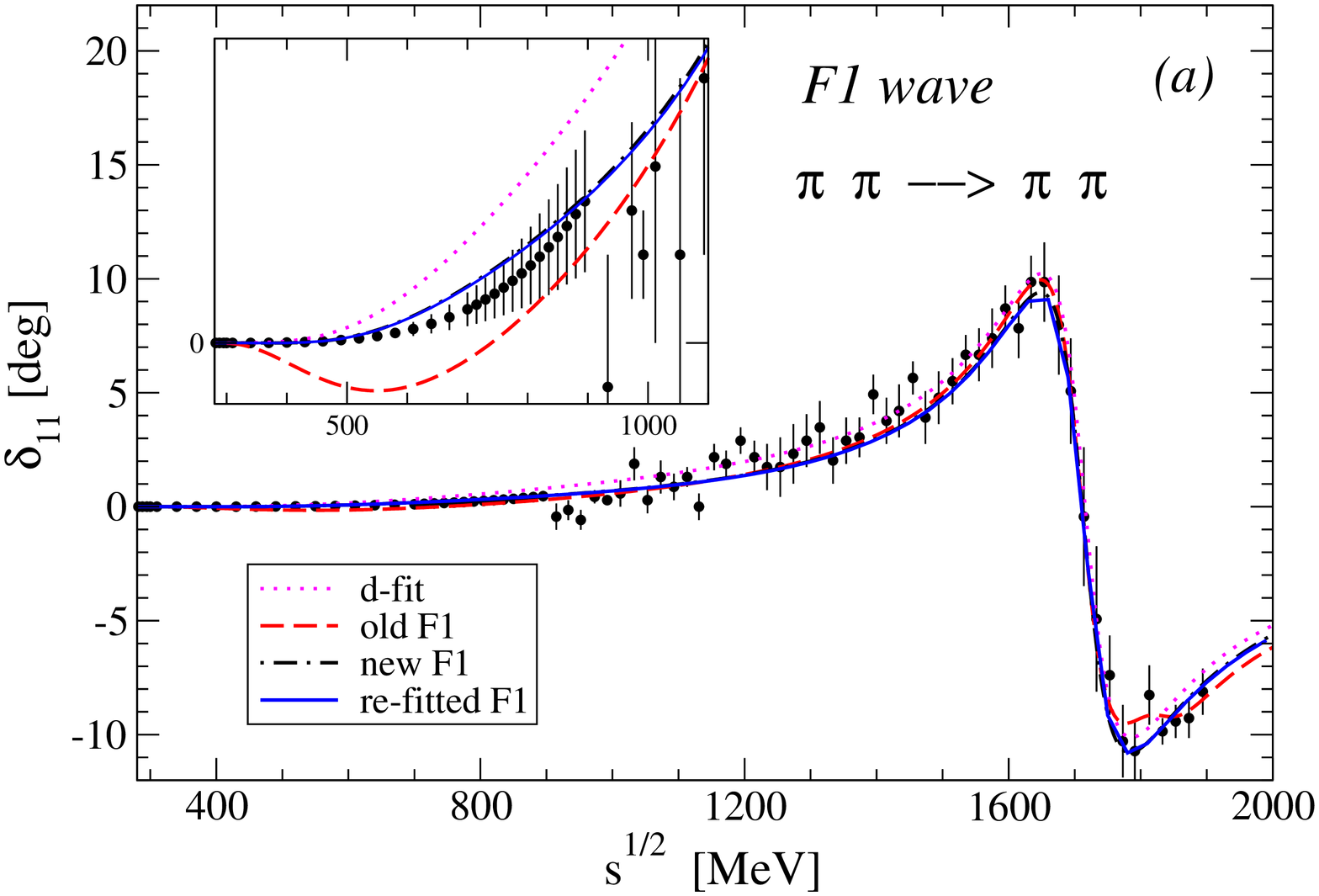}\\
\includegraphics[angle=270,scale=0.31]{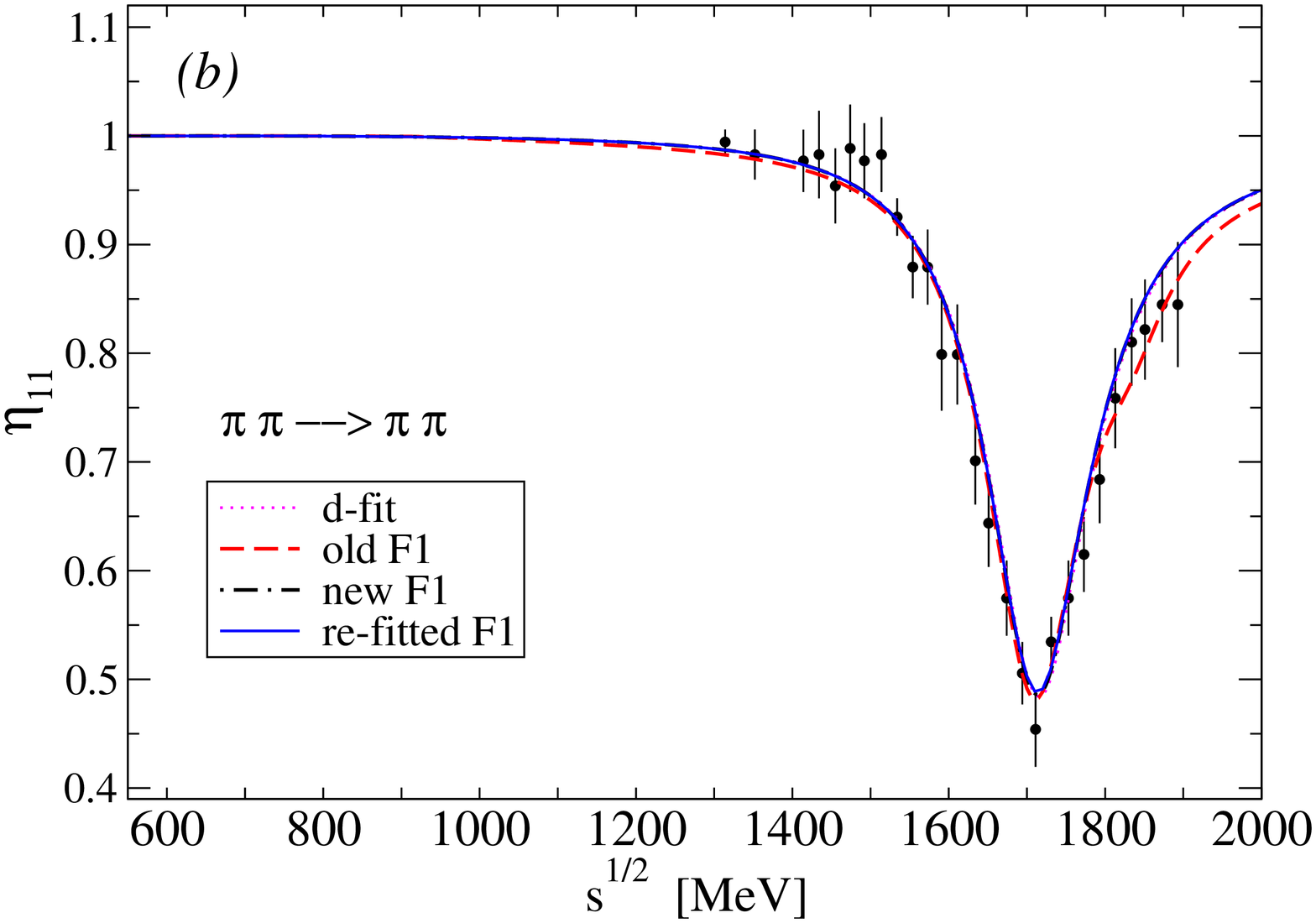}
\caption{Results of the New $F1$ (dash-dotted line), old $F1$~\cite{YuraDF} 
(dashed line), the fit only to the experimental data (dotted line) and refitted 
(after fitting to data and dispersion relations) $F1$ (solid line) amplitudes for 
the phase shift ($a$) and inelasticity parameter ($b$) of the $\pi\pi \to \pi\pi$ 
scattering are compared with experimental data from Ref.~\cite{Hyams} and 
the pseudo data (see the detail in ({\it a})).}
\label{figF1}
\end{figure}

To verify that the experimental data can be described purely by one state 
$\rho_3(1690)$ we fitted only the data with one Breit-Wigner form without background 
and assuming two channels 1 and 2 (so called ``d-fit"). 
The free parameters acquired similar values as 
those in the previous fit (Table~\ref{newF1}): $M_r=1715.9$, $f_{11}=294.0$, 
and $f_{12}=498.7$. The quality of the fit is a bit worse than of the previous one, 
$\chi^2/n.d.f.=110.18/(77-3)=1.49$ (compare the value of $\chi^2/{n.d.f.}$ 
without the pseudo data), but this fit clearly demonstrates that the resonance 
patterns revealed by both the phase shift and the inelasticity parameter are 
produced by single state $\rho_3(1690)$. The fit, however, overpredicts the 
pseudo data on $\delta_{11}$ in the threshold energy region (see the detail 
in Fig.~\ref{figF1}($a$)) and the calculated scattering length is by a factor 2 
larger than the value obtained from the GKPY equation~\cite{GKPY}.  

Results of the New $F1$ amplitude and of the simple fit only to the 
experimental data without background are compared with the data and 
with the old $F1$ amplitude~\cite{YuraDF} in Fig.~\ref{figF1}. 
The result of d-fit for $\delta_{11}$ is shifted upward with respect to the 
New $F1$ in the whole considered energy region. These results confirm necessity 
to introduce the background part of the amplitude (\ref{bgr_F1}) to describe 
correctly the pseudo data. Improvement is apparent also for energies above 
1.8~GeV. The resonance pattern seen in Fig.~\ref{figF1}($b$) is a little bit 
narrower for the New $F1$ amplitude than  for the old one.
%
%
\begin{table}
\begin{tabular}{ccccccc}  
\hline 
 state & PDG     &  $M_r$  & $f_{r1}$& $f_{r2}$& $f_{r3}$& $f_{r4}$ \\  
\hline 
$\rho_3(1690)$& 1688.8 $\pm$ 2.1 & 1714.0 & 293.1  & 498.3 & 0.0 &  0.0 \\
  \hline 
\end{tabular}
\caption{Parameters of the Breit-Wigner form (in MeV) for the New $F1$ amplitude. 
The mass of the resonance from PDG~\cite{PDG2014} is shown in the second 
column.} 
\label{newF1}
\end{table}

%
%

\section{Results of the dispersive analysis for the  \mbox{\boldmath $S$}-, 
\mbox{\boldmath $P$}-, \mbox{\boldmath $D$}-, and \mbox{\boldmath $F$}-wave  
\mbox{\boldmath $\pi\pi$}~amplitudes}

\subsection{Method of analysis}

The method of analysis is generally the same as in our previous 
work~\cite{BKN2014} which was briefly recalled in the Section~\ref{SPanalysis}.
As the initial amplitudes in the dispersive analysis we used   
the New $S0$-wave amplitude from~\cite{BKN2014} and the Old $P1$ amplitude  
from~\cite{Yura2010}, both with added polynomial near the threshold 
(the ``extended'' amplitudes in Ref.~\cite{BKN2014}). 
For the $D0$ and $F1$ waves we used the New amplitudes constructed in Sect. III and  
a correct behavior of the phase shift near the threshold was controlled 
by the pseudo data as in Sect. III. 
Since there are no resonances for $S2$ and $D2$ waves, the amplitudes for 
these isotensor waves were taken 
from~\cite{GKPY} and kept unchanged during the analysis. 
The analysis was performed with the same GKPY equations for the $S$ and $P$ 
waves as in Ref.~\cite{BKN2014} and with the GKPY equations for the $D$ and 
$F$ waves from the work~\cite{osdrdf}. 
In the modification of the amplitudes we allowed to change only those parameters 
that were expected to contribute significantly to the dispersive integrals, 
i.e. the low-energy resonances and the \pipi background. Definitions of the 
$\chi^2$ functions were analogous to Eqs.~(\ref{EqChi2Total}-\ref{EqChi2DR}). 

The analysis was split into three steps. In the first step, 
({\it the SP analysis}), only the $S0$ and $P1$ amplitudes were modified. 
In the fitting procedure, the matching energy and 
parameters of only $f_0(500)$ and $f_0(980)$ resonances and of 
the \pipi background were free in the $S0$ wave. 
Likewise in the $P1$ wave the matching energy and parameters of background and 
only the $\rho(770)$ resonance were fitted. Note that contrary 
to Ref.~\cite{BKN2014} the parameters of $f_0(1500)$ were not changed. 
Therefore the number of free parameters in the SP analysis 
decreased to 31. To make a comparison with the results from~\cite{BKN2014} 
possible we used in this step the phenomenological $D0$ and $F1$ 
amplitudes~\cite{GKPY} as in~\cite{BKN2014}. 

In the second step, ({\it the DF analysis}), we used the final $S0$ and $P1$ 
amplitudes from the SP analysis and the New $D0$ and $F1$ from Sect. III as 
initial amplitudes. 
In this step, only the latter two amplitudes were successively modified  
fitting the parameters of the Breit-Wigner forms of the $f_2(1270)$, 
$f_2(1525)$, $f_2(1640)$, $f_2(1810)$, $f_2(2125)$, and $f_2(2300)$ and 
$\rho_3(1690)$ resonances for the $D0$ and $F1$ partial-waves, respectively, 
together with the background parameters. The number of free parameters 
in the DF analysis was 31.    

In the third step, ({\it the SPDF analysis}), we started with the $S0$ and $P1$ 
amplitudes from the SP analysis and with the $D0$ and $F1$ amplitudes from 
the DF analysis. In this step we fitted again all free parameters considered 
above and arrived at the final form of all four amplitudes that we denote 
``re-fitted''. These amplitudes are optimized to the data and are consistent 
with the GKPY equations.

\subsection{Results of the \mbox{\boldmath SP} analysis}\label{SecresSP}

Values of the $\chi^2$ for each contribution in the $SP$ analysis are presented 
distinctly in Tables~\ref{TableChi2-S} and \ref{TableChi-DRsp}. The total $\chi^2/n.d.f.$ 
is $622.36/(572-31) = 1.15$ which is the same value as that obtained in 
Ref.~\cite{BKN2014} despite the fact that here we did not include the 
$f_0(1500)$ state. This justifies omitting parameters of $f_0(1500)$ form the set of 
free parameters and confirms a stability of the results. Comparing the $\chi^2/n.d.f.$ 
with the initial one ($\chi^2/n.d.f.=1520.3/541=2.81$) shows a big improvement in the fitting. 

The components of the $\chi^2_{Data}$ in the $S0$ wave have generally improved 
with a substantial change in the $\pi\pi$ $\rightarrow$ $\pi\pi$ and $K \bar K$ 
channels. 
Although the result in the $\pi\pi$ $\rightarrow$ $\eta \eta'$ channel is worse,  
the total $\chi^2_{Data}$ improves noticeably as in Ref.~\cite{BKN2014}. 
For the $P1$ wave, $\chi^2_{Data}$ of the ${\delta_{11}}$ slightly improves 
but it is slightly worse for ${\eta_{11}}$.  
As it was expected a substantial improvement was in the $\chi^2$ of the dispersion relations 
for $S0$ like in Ref.~\cite{BKN2014} which is due to the crossing symmetry. 
The $50\%$ improvement in the $\chi^2_{DR}$ for $P1$ is produced mainly by changes in 
the $S0$ wave which is coupled to $P1$ via the kernel in the GKPY equations~(\ref{EqGKPY}). 

%
%
\begin{table}[h!]
\hspace{.0cm}
\begin{tabular}{lcccccc}  
\hline
\hline
& \multicolumn{1}{c}{} & \multicolumn{1}{c}{} & \multicolumn{1}{c}{$S0$ wave} &
\multicolumn{1}{c}{} & \multicolumn{1}{c}{} & \multicolumn{1}{c}{} \\
\hline
& \multicolumn{1}{c}{$\chi^2_{Data}$} & \multicolumn{1}{c}{${\delta_{11}}$} & \multicolumn{1}{c}{${\eta_{11}}$} &
\multicolumn{1}{c}{${\delta_{12}}$} & \multicolumn{1}{c}{${|S_{12}|}$} & \multicolumn{1}{c}{${|S_{13}|}$} \\
\hline
initial  & 321.8  & 132.9 &  23.0 & 126.4 & 35.6 & 3.89  \\
re-fitted-SP & 282.9  &   118.8  &  19.4   &  118.1  &   21.3  &   5.40\\
\hline
& \multicolumn{1}{c}{} & \multicolumn{1}{c}{} & \multicolumn{1}{c}{$P1$ wave} &
\multicolumn{1}{c}{} & \multicolumn{1}{c}{} & \multicolumn{1}{c}{} \\
\hline

&& \multicolumn{1}{c}{$\chi^2_{Data}$} & \multicolumn{1}{c}{${\delta_{11}}$} & 
\multicolumn{1}{c}{${\eta_{11}}$} \\
\cline{2-5}
&initial  & ~302.8  & 264.1  &  38.7  \\  
&re-fitted-SP  & ~301.7  & 262.4  &  39.3  \\
\hline
\hline
\end{tabular}
\caption{Values of $\chi^2$ for data in the $S0$ and $P1$ waves before (initial) 
and after (re-fitted-SP) fitting in the SP analysis.}
\label{TableChi2-S}
\end{table}

%
%
\begin{table}[h!]
\begin{tabular}{lcccc}  
\hline
\hline
& \multicolumn{1}{c}{$\chi^2_{DR}$} & \multicolumn{1}{c}{$S0$} & \multicolumn{1}{c}{$S2$} &
\multicolumn{1}{c}{$P1$} \\
\hline
initial  & 895.7  & 842.9 &  8.43 & 44.3   \\
re-fitted-SP & 37.8  &   7.32  &  10.3   &  20.2 \\
\hline
\hline
\end{tabular}
\caption{Values of $\chi^2$ for the dispersion relations for the $S0$ and $P1$ 
waves  before (initial) and after (re-fitted-SP) fitting in the SP analysis.}
\label{TableChi-DRsp}   
\end{table}

\subsection{Results of the \mbox{\boldmath DF} analysis}\label{SecresDF}    

In order to improve agreement of the $D0$ and $F1$ wave amplitudes constructed in 
sections~\ref{fitsDwave} and~\ref{fitsFwave} with crossing symmetry condition, 
the New $D0$ and New $F1$ amplitudes have been fitted to the GKPY dispersion relations 
and to the data, where the $S0$- and $P1$-wave amplitudes were from the SP analysis 
(section~\ref{SecresSP}) and remained fixed. Hence, the total $\chi^2$ in 
Eq. (\ref{EqChi2Total}) was composed of eight parts: two parts for $\chi^2_{Data}(k)$ in 
Eq.~(\ref{EqChi2Data}) for $D0$ and $F1$ partial-waves and six parts for $\chi^2_{DR}(k)$ 
in Eq.~(\ref{EqChi2DR}) for all partial-waves, namely $S0$, $S2$, $P1$, $D0$, $D2$ and $F1$.

\subsection{Results of the \mbox{\boldmath SPDF} analysis}\label{SecFDR}

In the third-last step of our analysis, the total $\chi^2$ in Eq. (\ref{EqChi2Total}) was composed of 
ten parts. Four parts for $\chi^2_{Data}(k)$ in Eq. (\ref{EqChi2Data}) for $S0$, $P1$, $D0$,  
and $F1$ partial-waves and six parts for $\chi^2_{DR}(k)$ in Eq. (\ref{EqChi2DR}) for all 
considered partial-waves, namely $S0$, $S2$, $P1$, $D0$, $D2$, and $F1$. 
%
%
\begin{table}[h!]
\hspace{.0cm}
\begin{tabular}{lcccccc}  
\hline
\hline
\multicolumn{1}{c}{} & \multicolumn{6}{c}{$S0$ wave}\\
\hline
& \multicolumn{1}{c}{$\chi^2_{Data}$} & \multicolumn{1}{c}{$\delta_{11}$} & 
    \multicolumn{1}{c}{$\eta_{11}$} & \multicolumn{1}{c}{$\delta_{12}$} & 
     \multicolumn{1}{c}{$|S_{12}|$} & \multicolumn{1}{c}{$|S_{13}|$} \\
\hline
initial   & 321.8 & 132.9 & 23.0 & 126.4 & 35.6 & 3.89\\
re-fitted & 292.2 & 129.3 & 19.2 & 117.5 & 21.1 & 5.02\\
\hline
\multicolumn{1}{c}{} & \multicolumn{6}{c}{$P1$ wave}\\
\cline{2-5}
 \multicolumn{2}{c}{} & \multicolumn{1}{c}{$\chi^2_{Data}$} & \multicolumn{1}{c}{$\delta_{11}$} & 
\multicolumn{1}{c}{$\eta_{11}$} &  \multicolumn{2}{c}{} \\
\cline{2-5}
& \multicolumn{1}{l}{initial}    & ~302.8  & 264.1  &  38.7 & \multicolumn{2}{c}{}\\  
& \multicolumn{1}{l}{re-fitted}  & ~299.3  & 260.7  &  38.6 & \multicolumn{2}{c}{}\\
\hline
\hline
\end{tabular}
\caption{Values of $\chi^2$ for data in the $S0$ and $P1$ waves before (initial) 
and after (re-fitted) fitting in the SPDF analysis.}
\label{TableChi2-S-SPDF}
\end{table}

%
%
\begin{table}[h!] 
\centering
\begin{tabular}{lccccccccc}  
\hline
\hline
\multicolumn{1}{c}{}& \multicolumn{5}{c}{$D0$ wave} & \multicolumn{1}{c}{ } & 
\multicolumn{3}{c}{$F1$ wave}\\
\cline{2-6} \cline{8-10} 
\multicolumn{1}{c}{}& \multicolumn{1}{c}{$\chi^2_{Data}$} & \multicolumn{1}{c}{${\delta_{11}}$} & \multicolumn{1}{c}{${\eta_{11}}$} &
 \multicolumn{1}{c}{${|S_{13}|}$} & \multicolumn{1}{c}{${|S_{14}|}$} & \multicolumn{1}{c}{}  &\multicolumn{1}{c}{$\chi^2_{Data}$} & \multicolumn{1}{c}{${\delta_{11}}$} & 
\multicolumn{1}{c}{${\eta_{11}}$} \\
\cline{2-6}\cline{8-10}
initial   & 242.3 & 137.0 &  76.7 & 21.1 & 7.44 & & 136.5  & 120.4  &  16.1\\
re-fitted & 218.7 & 128.3 &  64.1 & 18.1 & 8.1  & & 137.3  & 120.8  &  16.6\\
\hline
\hline
\end{tabular}
\caption {Values of $\chi^2$ for data in the $D0$ and $F1$ waves before (initial) 
and after (re-fitted) fitting in the SPDF analysis.}
\label{TableChi2-DataDF}
\end{table}

%
%
\begin{table}[h!]
\begin{tabular}{lccccccc}  
\hline
\hline
& \multicolumn{1}{c}{$\chi^2_{DR}$} & \multicolumn{1}{c}{$S0$} & \multicolumn{1}{c}{$S2$} &\multicolumn{1}{c}{$P1$}& \multicolumn{1}{c}{$D0$} & \multicolumn{1}{c}{$D2$} &
\multicolumn{1}{c}{$F1$} \\
\hline
initial  & 313.2 & 107.8 &  24.4 & 18.0 & 125.1 &  16.8 & 21.2  \\
re-fitted & 113.9 & 5.05 &  17.9 & 27.8 & 27.5  &  15.8   &  19.9 \\
\hline
\hline
\end{tabular}
\caption{Values of $\chi^2$ for the dispersion relations for all waves before (initial) and after (re-fitted) fitting in the SPDF analysis.}
\label{TableChi-DRspdf}
\end{table}

The value of the $\chi^2$ in the SPDF analysis after fitting to the data and dispersion relations 
is $\chi^2/n.d.f.=1061.5/895=1.19$ which is almost the same as the value in the SP analysis 
though it includes also contributions of $\chi^2_{Data}$ from the multi-channel $D0$ and 
$F1$ amplitudes. Remind that in the SP analysis we used phenomenological  
parameterizations of the $D0$- and $F1$-wave $\pi\pi$ amplitudes from \cite{GKPY}.

Tables \ref{TableChi2-S-SPDF}-\ref{TableChi-DRspdf} show the $\chi^2$ of the data 
and the dispersion relations for all waves distinctly. The values of the $\chi^2_{Data}$ 
for the $S0$ and $D0$ waves have generally improved except in the $\pi\pi$ $\rightarrow$ 
$\eta\eta'$ channel for the $S0$ wave and $\pi\pi$ $\rightarrow$ 
$\eta\eta$ channel for the $D0$ one. Description of data in this inelastic channel tends to be worse for 
both $S0$ and $D0$ waves (see ${|S_{13}|}$ in Table~\ref{TableChi2-S-SPDF} 
and ${|S_{14}|}$ in Table~\ref{TableChi2-DataDF}) which can be attributed to the coupling 
between these two waves.
Description of data in the $P1$ wave slightly improved for both phase shift and inelasticity.
The $\chi^2_{Data}$ for the $F1$ wave almost did not change because the parameters of 
the New $F1$ amplitude are not too much affected by fitting the dispersion relations. 
All components of the $\chi^2_{DR}$, 
except  that
for $P1$, are smaller after fitting with a substantial improvement for the $S0$ and $D0$ waves. 

Note that the initial values of the $\chi^2_{DR}$ for the $S0$ and $P1$ waves in 
Table~\ref{TableChi-DRspdf} differ from the final values in Table~\ref{TableChi-DRsp} 
due to different $D0$ and $F1$ amplitudes used in the analysis.
In the SP analysis, the $D0$ and $F1$ amplitudes are from~\cite{GKPY} and not modified 
while in the SPDF analysis we used the New $D0$ and New $F1$ amplitudes. 
It is especially well seen for the initial value of the $\chi^2_{DR}$ for the $P1$ wave 
which is smaller in the SPDF analysis (18.0) than in the SP one (44.3).
Finally it becomes larger but still comparable with the final value in the SP analysis. 
This suggests quite strong influence of the other amplitudes (especially New $F1$) 
on the $P1$ amplitude in the dispersive analysis.

Comparing results of the SPDF and SP analysis for the $S0$ wave we see that description 
of the phase shift $\delta_{11}$ is slightly worse in the SPDF analysis. This we attribute to the 
influence of the New $D0$ and New $F1$ amplitudes, especially of the former as there is 
stronger correlation between these two waves ($S0$ and $D0$). This can be particularly 
well seen in a comparison of the final (re-fitted-SP) values of $\chi^2_{DR}$ for $S0$ and $S2$ in 
Table~\ref{TableChi-DRsp} with the initial values in Table~\ref{TableChi-DRspdf}. 

In Table \ref{TableChi2-f1500} we show influence of fitting parameters of the $f_0(1500)$ 
on the results in the SP and SPDF analyses to see significance of this resonance in the analysis.
As it was expected from the previous analysis in Ref.~\cite{BKN2014} for the $S$ and $P$ waves, 
results are not too much sensitive to changes of the  $f_0(1500)$-resonance parameters.   
Although the $\chi^2$ is smaller when parameters of $f_0(1500)$ resonance are free, 
the $\chi^2/n.d.f.$ do not change. 
This corroborates our previous results of the analysis in Ref.~\cite{BKN2014} where parameters 
of $f_0(1500)$ changed very slightly.
 
%
%
\begin{table}[h!]
\hspace{.0cm}
\begin{tabular}{lcc}  
\hline
\hline
& \multicolumn{1}{c}{$\chi^2$/$n.d.f.$ ($f_0(1500)$ fixed) } & 
  \multicolumn{1}{c}{$\chi^2$/$n.d.f.$ ($f_0(1500)$ free)}  \\
\hline
$\chi^2$(SP)  & 622.4/541=1.15 & 609.7/529=1.15   \\   
$\chi^2$(SPDF)   & 1061.5/895=1.19 & 1053.6/883=1.19   \\ 
\hline
\hline
\end{tabular}
\caption{Values of the $\chi^2$/$n.d.f.$ for the SP and SPDF analyzes when parameters 
of the $f_0(1500)$ resonance are fixed or free in fitting.}
\label{TableChi2-f1500}
\end{table}

\section{Discussion}
In Tables \ref{par_poles_S0-final} - \ref{par_bgr_P1-final}, we provide a comparison of 
parameters of the original (``New" amplitude from \cite{BKN2014} in case of the $S0$ wave) 
and re-fitted $S0$- and $P1$-wave amplitudes in the SPDF analysis. 
A substantial change of the parameters after the analysis is a shift of the position of 
the $\sigma$ pole on sheet II. The new pole position, ($477.6 \pm 14 -i\,302.0 \pm 14$~MeV) 
is within three standard deviations consistent with the result in Ref.~\cite{BKN2014}, 
($445.2 \pm 14 - i\,296.4 \pm 14$~MeV) demonstrating a stability of the results. 
This new value is also compatible with that presented by Particle Data Group~\cite{PDG2014} 
$(400-550) -i(200-350)$~MeV.  
The pole positions of the $f_0(980)$ resonance did not change too much. 
The real and imaginary parts became only slightly smaller and are compatible with the values 
presented in~\cite{BKN2014}. 
In the $P1$ wave, the position of the $\rho(770)$-pole on the Riemann sheet VI was shifted 
significantly toward smaller energies and closer to the real axis. This, however, does not affect 
a data description as this pole is far from the physical region. Note, however, that the sheet VI 
is directly connected with the physical one above the $\rho\sigma$ threshold and therefore 
a pole lying on the sheet VI near the real axis above the $\rho\sigma$ threshold, in many cases, 
can influence appreciably description of the data.
  
The background parameters of the $S0$ and $P1$ waves changed moderately showing that 
the background part plays only a marginal role in the amplitude. 
The matching energy $\sqrt{s_{00}}$ became bigger in comparison to that in~\cite{BKN2014}. 
One may attribute this to the influence of the New $D0$ wave. On the contrary the matching 
energy $\sqrt{s_{01}}$ for the $P1$ wave is similar to that in~\cite{BKN2014}. 
For the $P1$ wave the background parameter $b$ acquired a very small negative value which 
affects inelasticity of the $P1$ wave, slightly violating unitarity for energies near 1.8~GeV 
(see  Fig.~\ref{figP1-1}(b)). 
%
%
\begin{table}[h!]
\begin{tabular}{crrr}
\hline
\hline
Sheet  &           & ~~~~~~original      & ~~re-fitted\\
\hline
 & \multicolumn{3}{c}{$f_0(500)$} \\   
II  & $E_r$        &   $562.9 $ & $477.6$\\ 
    & $\Gamma_r/2$ &   $417.1  $ & $302.0$\\ 
III & $E_r$        &   $594.7 $ & $717.6$\\ 
    & $\Gamma_r/2$ &  $417.1  $ & $300.4$\\ 
VI  & $E_r$        &   $615.1 $ & $422.2$\\ 
    & $\Gamma_r/2$ &   $417.1  $ & $448.5$\\ 
VII & $E_r$        &   $583.3 $ & $602.7$\\ 
    & $\Gamma_r/2$ &   $417.1  $ & $206.7$\\ 
\hline
 & \multicolumn{3}{c}{$f_0(980)$}\\
II  & $E_r$        &  $1007.6 $  & $999.6 $ \\ 
    & $\Gamma_r/2$ &  $29.4 $    & $\,\,\, 21.0 $ \\ 
III & $E_r$        &  $984.5 $   & $ 975.5$ \\ 
    & $\Gamma_r/2$ &  $55.1 $    & $\,\,\, 22.1 $ \\ 
\hline
\hline
\end{tabular}
\caption{Real ($E_r$) and imaginary ($\Gamma_r/2$) parts of poles 
on the Riemann sheets of two lowest resonances in the $S0$ 
amplitude before (original) and after (re-fitted) the full analysis. 
The values are in MeV.}
\label{par_poles_S0-final}
\end{table}

%
%
\begin{table}[h!]
\hspace{.0cm}
\begin{tabular}{ccc}  
\hline
\hline
Parameter    & ~~original~~  & ~~re-fitted~~\\
\hline
$a_{11}$      &   -0.0131 &-0.07870 \\
$a_{1\sigma}$ &  0.0    & 0.13210 \\
$a_{1v}$      &  0.046 & -0.13380  \\
$a_{1\eta}$   & -0.0302 & -0.01832  \\
$b_{1\sigma}$ &  0.0    &0.09207 \\
$b_{1v}$      & 0.0573 & 0.01855  \\
$b_{1\eta}$   &  0.0    & -0.03753 \\
$\sqrt{s_{00}}$      &   406.5  & 495.0 \\
\hline
\hline
 \end{tabular}
\caption{Values of the background parameters and the matching energy 
$\sqrt{s_{00}}$ (in MeV) for the $S0$ wave before (original) and after 
(re-fitted) the full analysis.}
\label{par_bgr_S0-final}
\end{table}

%
%
\begin{table}[h]
\begin{tabular}{ccrr}
\hline
\hline
Sheet  &           & ~~~~~~original      & ~~re-fitted\\
\hline
 & \multicolumn{3}{c}{$\rho(770)$}\\  
II  & $E_r$        &  $766.0 $   & $765.4 $ \\ 
    & $\Gamma_r/2$ &  $72.5  $   & $73.0  $ \\ 
III & $E_r$        &  $758.7 $   & $ 799.4$ \\ 
    & $\Gamma_r/2$ &  $72.5  $   & $ 53.7 $ \\ 
VI  & $E_r$        &  $753.5 $   & $1.28$ \\ 
    & $\Gamma_r/2$ &  $72.5   $   & $ 0.49$ \\ 
VII & $E_r$        &  $760.8.2 $   & $ 1051.5$ \\ 
    & $\Gamma_r/2$ &  $72.5  $   & $ 8.09 $ \\
\hline
\hline
\end{tabular}
\caption{The same as in Table~\ref{par_poles_S0-final} but for the $\rho(770)$ 
resonance in the $P1$ wave.}
\label{par_poles_P1-final}
\end{table}

%
%
\begin{table}[h!]
\hspace{.0cm}
\begin{tabular}{ccc}  
\hline
\hline
Parameter    & ~~original~~  & ~~re-fitted~~\\
\hline
$a$      & -0.2860 & -0.33148 \\
$b$      & 0.00012 &  -0.00008\\
$\sqrt{s_{01}}$ &  643.6  &  637.3 \\
\hline
\hline
 \end{tabular}
\caption{The same as in Table~\ref{par_bgr_S0-final} but for the $P1$ wave.}
\label{par_bgr_P1-final}
\end{table}

\begin{figure}[h!]
\includegraphics[angle=270,scale=0.31]{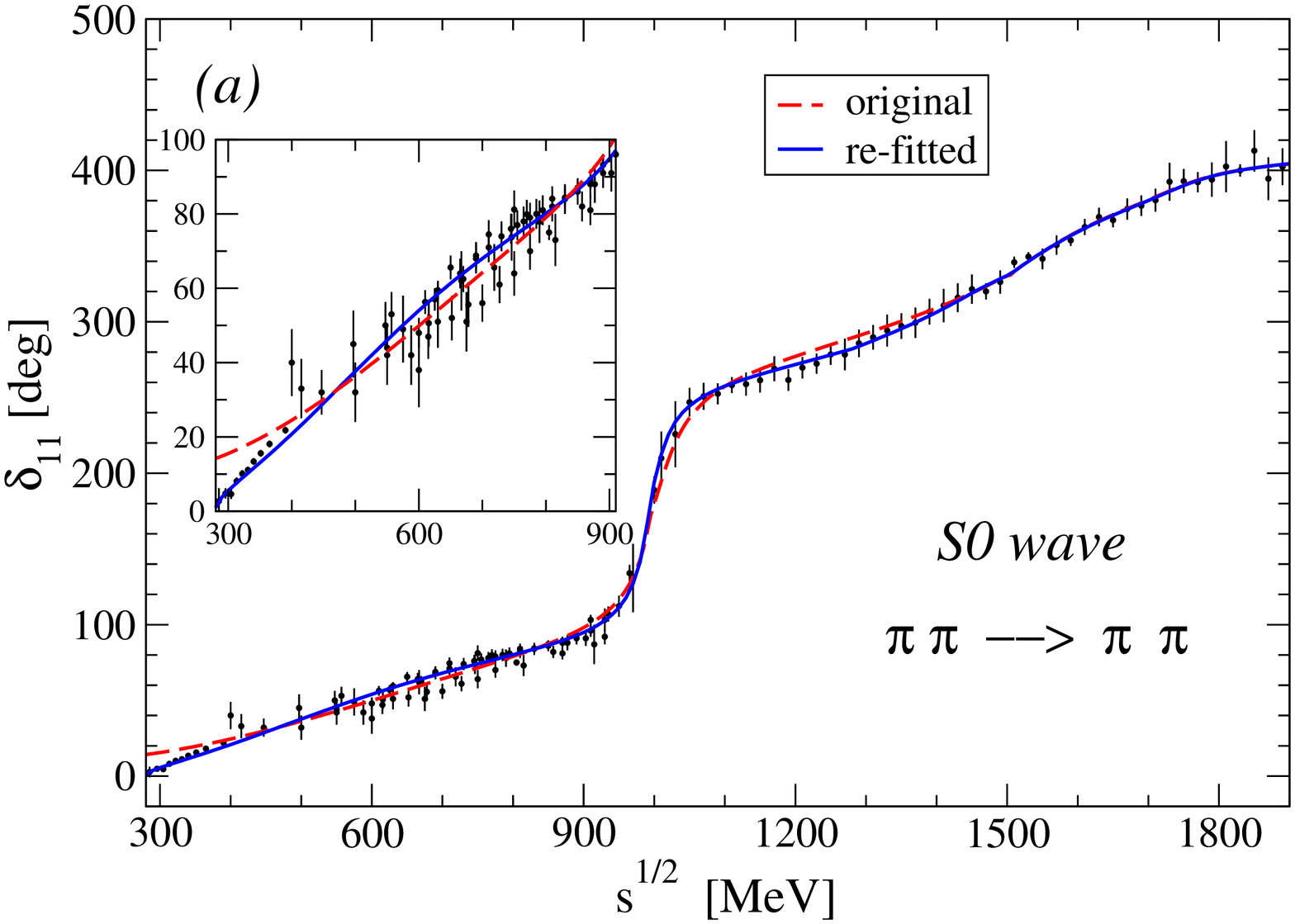}\\
\includegraphics[angle=270,scale=0.31]{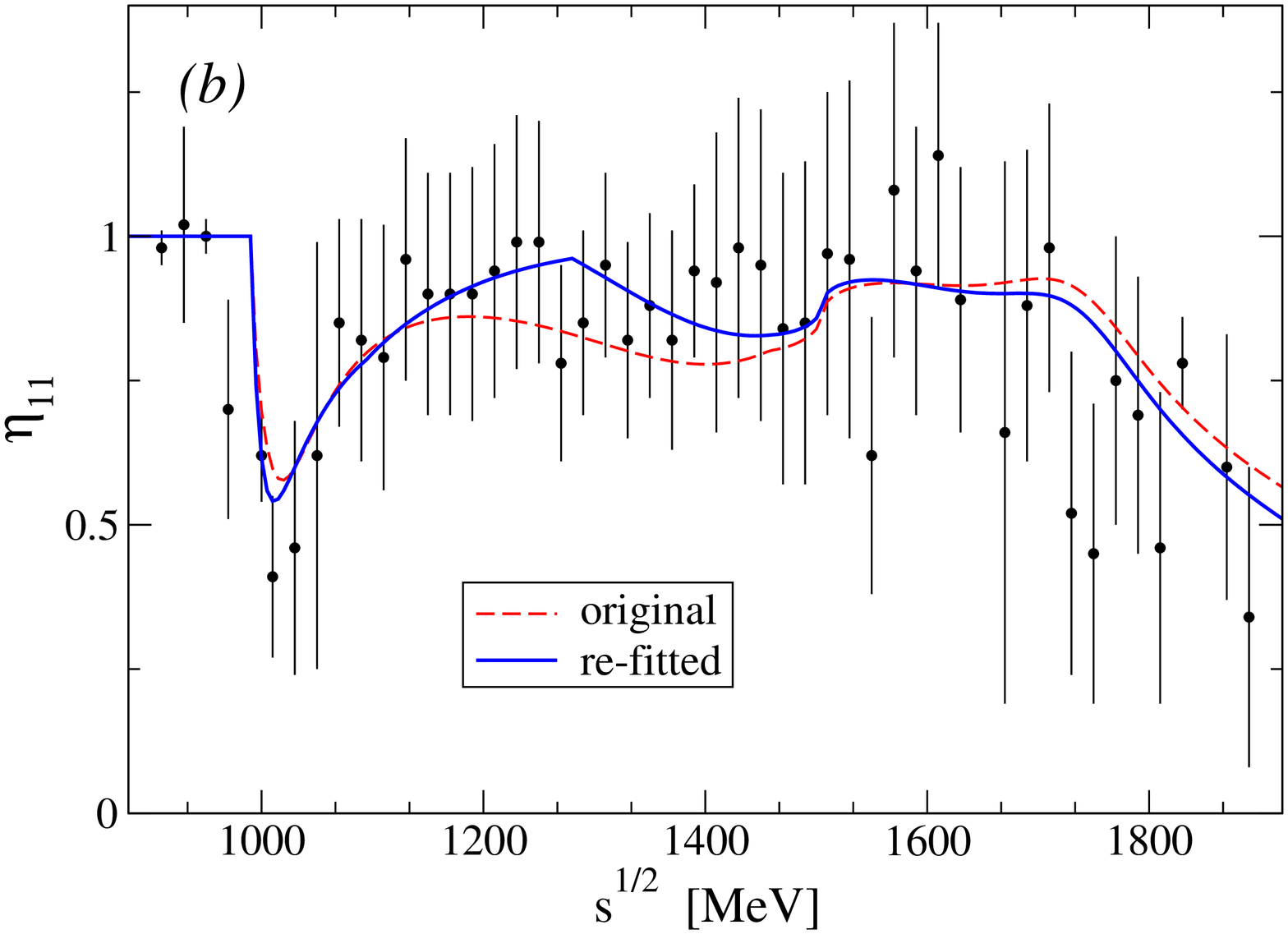}
\caption{Results of the original $S0$ (dashed line) and re-fitted (after fitting to data and dispersion relations) $S0$ (solid line) amplitudes for the phase shift ($a$) and inelasticity ($b$) 
of the $\pi\pi \to \pi\pi$ scattering are compared with experimental data.}
\label{figS0-1}
\end{figure}

\begin{figure}[h!]
\includegraphics[angle=270,scale=0.31]{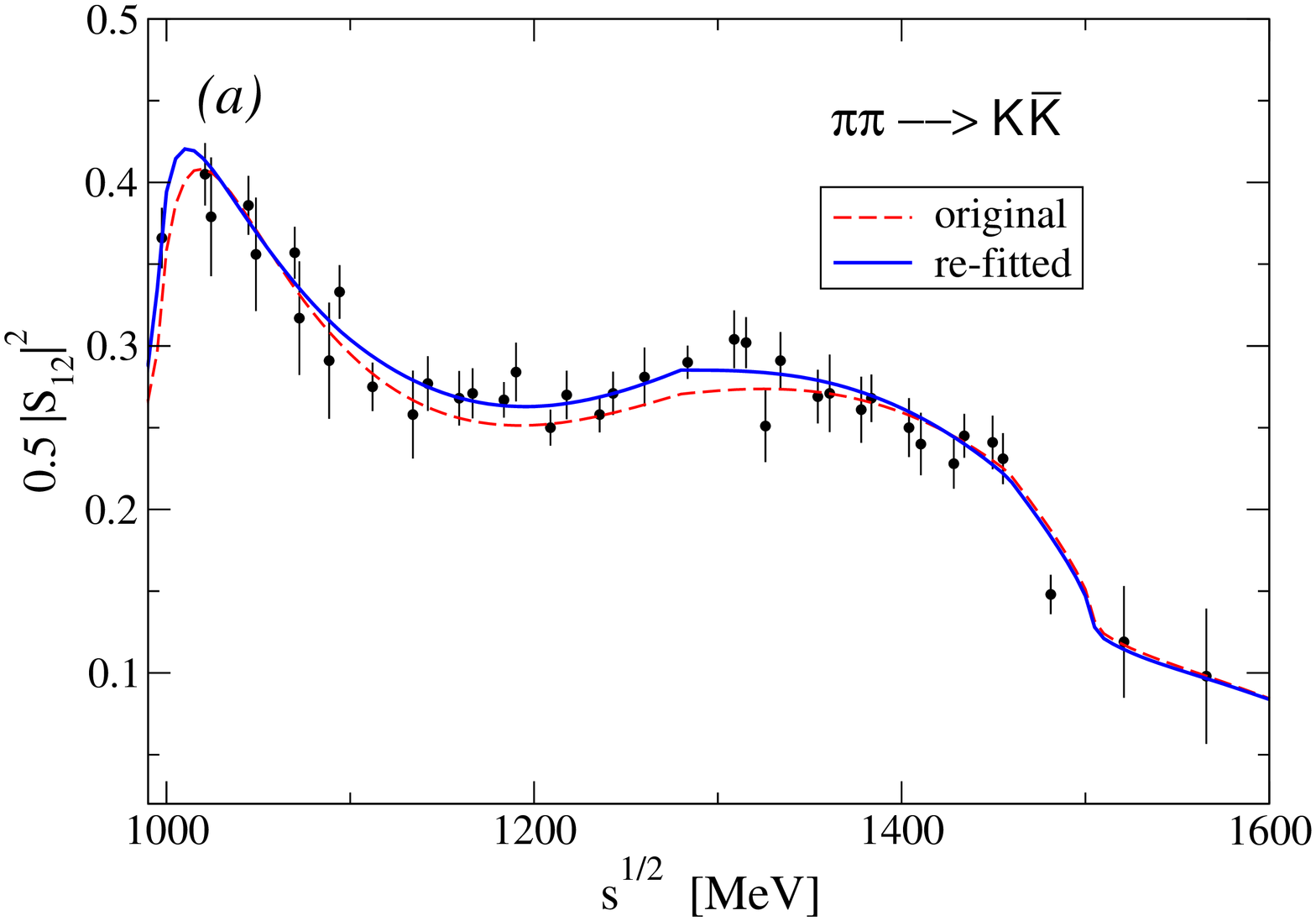}\\
\includegraphics[angle=270,scale=0.31]{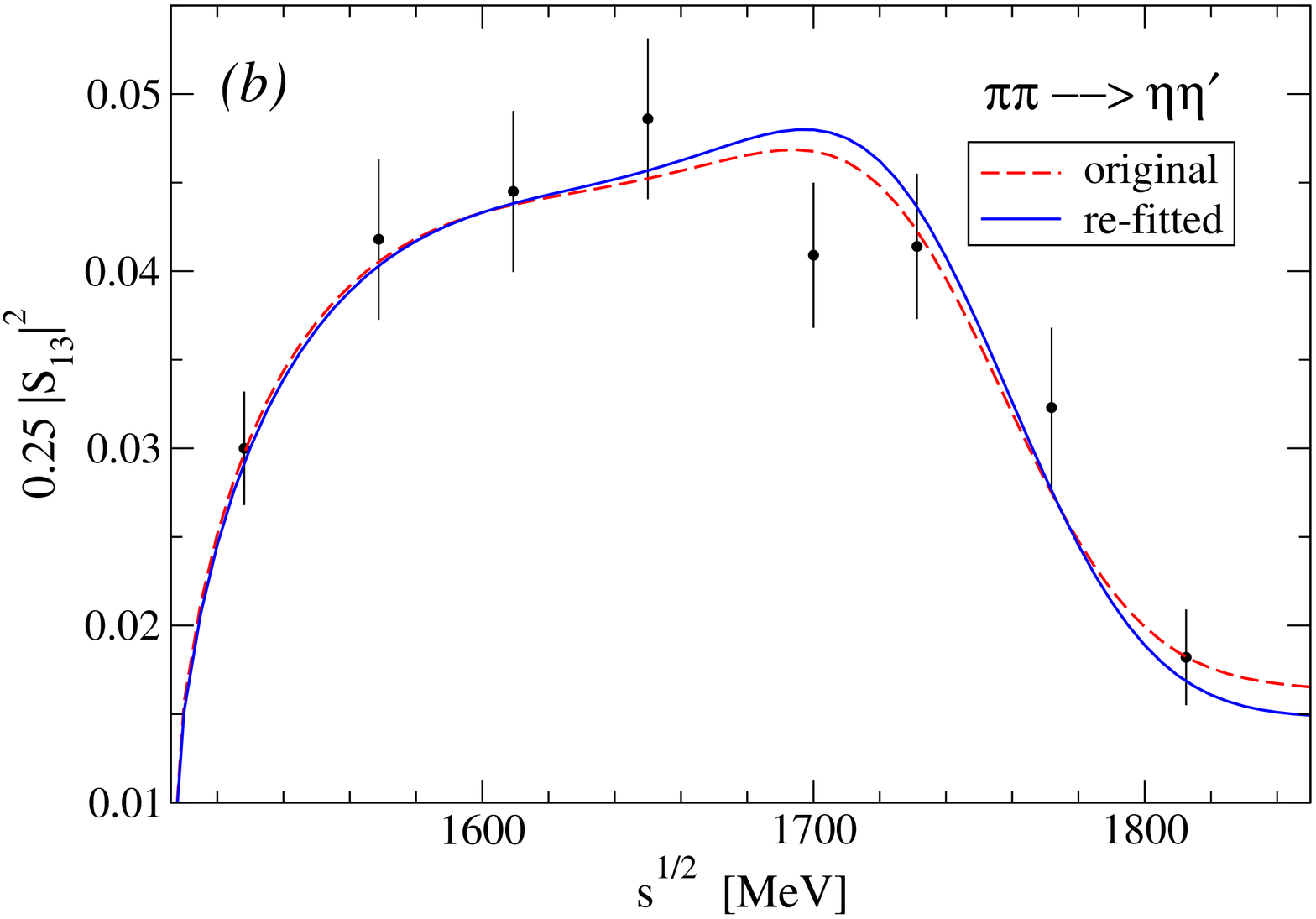}
\caption{The same as in Fig.~\ref{figS0-1} for the squared modulus of the S matrix 
for the $\pi\pi \to K\bar{K}$ ($a$) and $\pi\pi \to \eta\eta'$ ($b$).}
\label{figS0-2}
\end{figure}

\begin{figure}[h!]
\includegraphics[angle=270,scale=0.31]{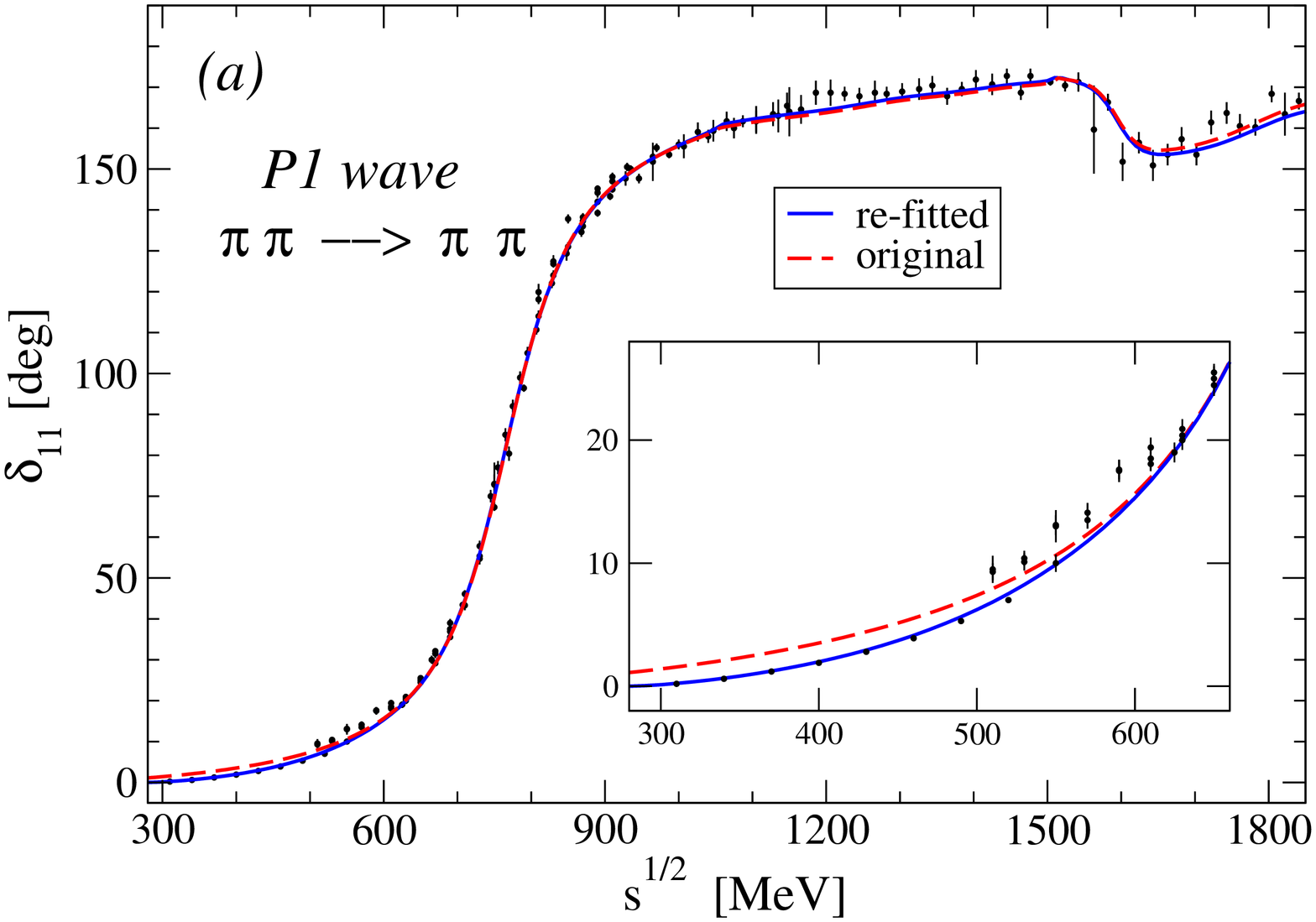}\\
\includegraphics[angle=270,scale=0.31]{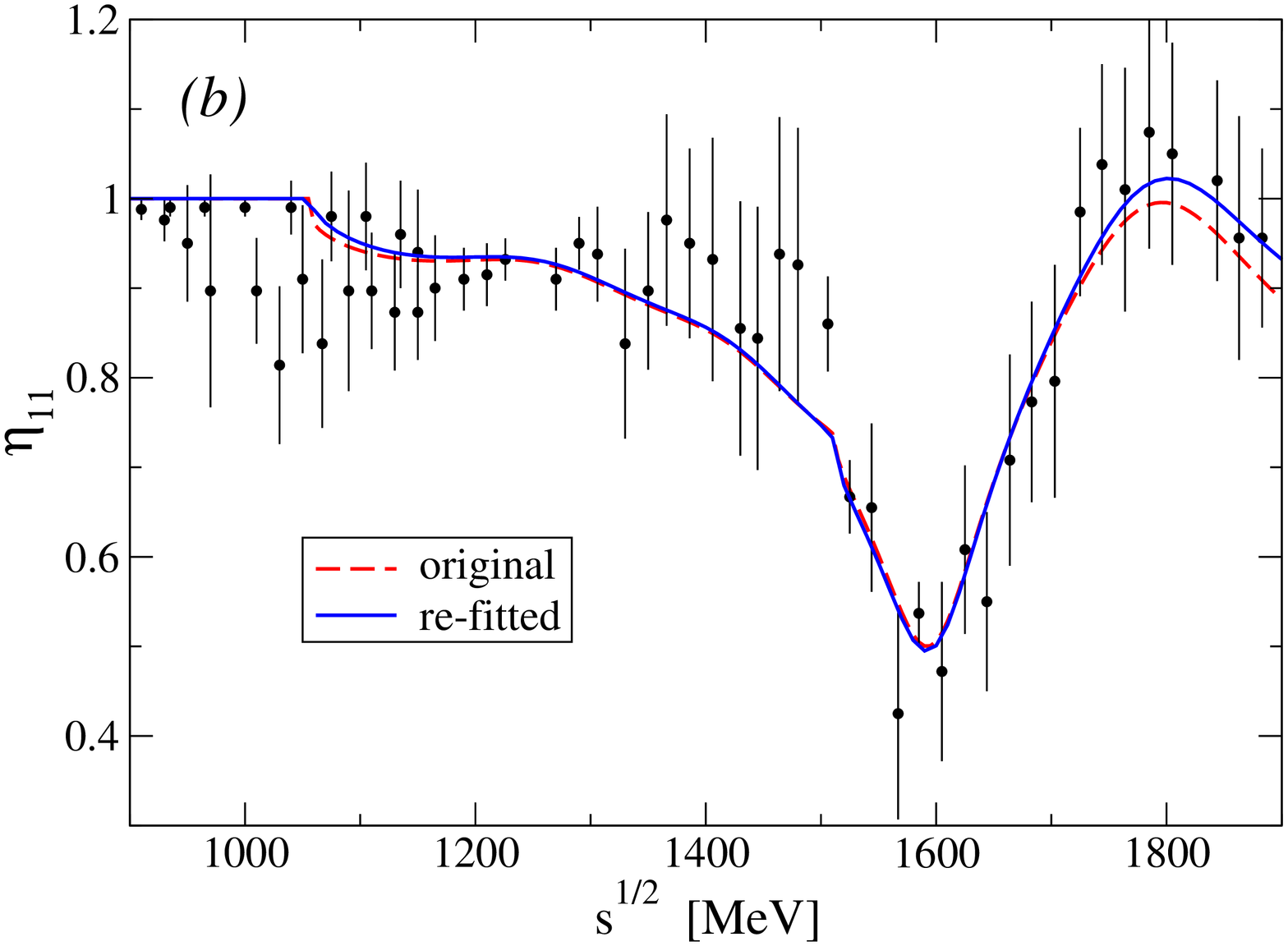}
\caption{Results of the original $P1$ (dashed line) and re-fitted (after fitting 
to data and dispersion relations) $S0$ (solid line) amplitudes for the phase shift 
($a$) and inelasticity ($b$) 
of the $\pi\pi \to \pi\pi$ scattering are compared with experimental data
and pseudo data points.}
\label{figP1-1}
\end{figure}

Figures~\ref{figS0-1} - \ref{figP1-1} illustrate the results of the re-fitted $S0$ and $P1$ 
amplitudes for the phase shift and inelasticities in the $\pi\pi$ $\rightarrow$ $\pi\pi$, 
$K \bar K$ and $\eta\eta'$ channels compared to the original amplitudes and available 
experimental data. Our final results describe the data very well in all considered channels. 
An improvement is especially apparent for the elastic phases $\delta_{11}$ for both waves 
in the low-energy region. 
The turn observed at 1.28~GeV in Fig.~\ref{figS0-1}(b) for inelasticity of the $S0$ wave in 
$\pi\pi$ $\rightarrow$ $\pi\pi$ is due to opening of the $\sigma \sigma$ 
channel included in the background part~\cite{BKN2014}. 
%
%
%
%
%
%

The new (re-fitted) parameters of the $D0$ and $F1$ waves in the SPDF analysis are given 
in Tables \ref{newD0-fitted} and \ref{newF1-fitted}. The re-fitted background parameters 
of the $D0$ amplitude are: $\alpha_{11}= 0.0011853$, $\alpha_{13}=0.037747$, 
$\alpha_{10}=-0.46722$, $\beta_1 = 0.15631$, $\beta_3 = -8.5280$, $\beta_4 = -11.4446$, 
$\gamma_1 = -0.31272$, $\gamma_3 = 9.9804$, and $\gamma_4 =14.8899$. 
The re-fitted background parameters of the $F1$ amplitude are: $a_\alpha = 0.0000132$, 
$a_\beta = -0.00102$, and $a_\gamma = 0.0000151$.
Note that contributions of the big magnitudes of the parameters $\beta_3$ and $\gamma_3$ 
in the $K\bar{K}$ channel and $\beta_4$ and $\gamma_4$ 
in the $\eta\eta$ channel tend to cancel each other above the effective vector-vector channel 
($s_v$) in the background part, see the formulas and text below Eq.~(\ref{JostDet_bgrD}). 
Similar correlations between $\beta_j$ and $\gamma_j$ were observed also in the data analysis 
performed in Sect. IIIB.

A comparison of Tables \ref{newF1} and \ref{newF1-fitted} shows that the parameters of the 
$F1$ wave changed only slightly as it can be also seen from a comparison of the initial and 
re-fitted values of the $\chi^2_{Data}$ in Table~\ref{TableChi2-DataDF}. 
Such comparison is more complicated for the $D0$ amplitude since there are more resonance 
states with more free parameters. 
Analysis shows that some $D0$ resonances are irrelevant 
and some of them have weighty 
decay widths ($f_{ri}$) only in two or three channels. 
For example, the decay widths of $f_2(1640)$ into the $\pi\pi$ and $\eta\eta$ channels 
are practically zero after the dispersive analysis (compare Tables~\ref{newD0} and 
\ref{newD0-fitted}) 
which is compatible with PDG~\cite{PDG2014} where only the $2\pi 2\pi$ and $K \bar K$ decays are seen.
Similarly the $f_2(2300)$ resonance 
reveals a very weak coupling/branching to the $2\pi 2\pi$ channel after the 
dispersive analysis. The $f_2(1525)$ resonance with a negligible decay width into the $\pi\pi$ 
channel~\cite{PDG2014} has also a small value of the parameter $f_{r1}$ in our analysis. 
Note that in the latest issue of PDG~\cite{PDG2014} only decays of the $f_2(1270)$ and $f_2(1525)$ 
resonances into the $\pi\pi$, $2\pi 2\pi$, $K \bar K$, and $\eta\eta$ channels are 
precisely determined  whereas for the other resonance states mostly a status ``seen'' 
is reported.
%
%
\begin{table}[h]
\begin{tabular}{ccccccc}  
\hline 
\hline
 state &   $M_r$  & $f_{r1}$& $f_{r2}$& $f_{r3}$& $f_{r4}$ \\  
\hline 
$f_2(1270)$&  1275.5 & 451.6  & 77.3 & 121.0 &  64.7 \\
$f_2(1525)$&  1570.7 &  76.3  & 400.8  & 999.0 & 922.9 \\ 
$f_2(1640)$&  1639.0 &  0.002  & 356.3 & 222.0 &  0.0002 \\
$f_2(1810)$&  1815.0 &  120.4  & 813.0 & 999.0 &  873.2 \\
$f_2(2150)$&  2157.0 & 11.7  & 842.8 & 803.4 & 999.0 \\
$f_2(2300)$&  2181.6 & 27.0  & 0.0005 & 453.5 & 142.9 \\
  \hline 
  \hline
\end{tabular}
\caption{Parameters of the Breit-Wigner form (in MeV) for the $D0$ 
wave after fitting in the SPDF analysis. The initial values of the 
parameters are in Table~\ref{newD0}. The masses were not changed in 
the dispersive analysis.} 
\label{newD0-fitted}
\end{table}

%
%
\begin{table}[h]
\begin{tabular}{ccccccc}  
\hline
\hline 
 state &   $M_r$  & $f_{r1}$& $f_{r2}$& $f_{r3}$& $f_{r4}$ \\  
\hline 
$\rho_3(1690)$& 1713.7 & 291.6  & 497.8 & 0.0 &  0.0 \\
  \hline 
  \hline
\end{tabular}
\caption{Parameters of the Breit-Wigner form (in MeV) for the 
$F1$ wave after fitting in the SPDF analysis. The initial values 
of the parameters are in Table~\ref{newF1}. The mass of the resonance 
was not changed in the dispersive analysis.}
\label{newF1-fitted}
\end{table}

Results for the phase shifts and inelasticities for the $D0$ and $F1$ waves are presented 
in Figs.~\ref{figD0-1} - \ref{figF1}. 
The dispersive analysis practically did not affect the description of the elastic phase shifts 
$\delta_{11}$ in both waves. The same holds true for the inelasticity parameter in the $F1$ wave. 
On the contrary inelasticity in the $D0$ wave for the $\pi\pi \to \pi\pi$ channel has improved significantly, especially around 
1 and 1.3~GeV. In the 1~GeV region the New $D0$ amplitude slightly violates unitarity which 
has been cured in its re-fitted version. The description becomes slightly worse around 1.5~GeV. 
One can also say that the modulus squared of $S_{13}$ ($\pi\pi$ $\rightarrow$ $K \bar K$) 
is also more realistic after the dispersive analysis, especially in the peak region. 
Noticeable changes are also apparent in behavior of $|S_{14}|^2$ 
($\pi\pi$ $\rightarrow$ $\eta\eta$) in Fig.~\ref{figD0-2}($b$). 

We have performed various fits to learn which $D0$ resonance states are dominant or 
ineffective in our approach. Results show that not all first four lightest states,  
i.e. $f_2(1270)$, $f_2(1430)$, $f_2(1525)$, and $f_2(1640)$, are the most relevant 
states in the analysis, as one would naively expect. 
This is demonstrated in Table \ref{TableChi2-DN} which shows the $\chi^2$ for the fits 
with one particular resonance omitted. 

The $\chi^2$ was composed of the $\chi^2_{Data}$ 
for the $D0$ wave and $\chi^2_{DR}$ for all partial-waves.  
Values of the $\chi^2$ in the fits No. 9, 11, and 12 are almost equal to the value in 
the fit No.~1 with all resonance states included. Accordingly, one can conclude that 
the $f_2(2010)$, $f_2(2300)$, and $f_2(2340)$ resonances are insignificant 
in the description.
Values of the $\chi^2$ in the fits No. 3, 5, 7, and 8 show that although the 
$f_2(1430)$, $f_2(1640)$ , $f_2(1910)$ and $f_2(1950)$ do not play a very important 
role in the data description one should keep them. 
The other states, $f_2(1525)$, $f_2(1810)$, and $f_2(2150)$, influence behavior 
of the amplitude even more. Obviously the most significant resonance which dominates 
behavior of the amplitude is the $f_2(1270)$ resonance.
%
%
\begin{table}[h!]
\centering
\begin{tabular}{ccccc}  
\hline
\hline
\multicolumn{1}{c}{Fit No.} & \multicolumn{1}{c}{} & \multicolumn{1}{c}{$\chi^2/n.d.f.$} \\
\hline
\multicolumn{1}{c}{1} & \multicolumn{1}{c}{ All states included} & \multicolumn{1}{c}{326.21/330=0.9885} \\
\hline
\multicolumn{1}{c}{} & \multicolumn{1}{c}{Omitted Resonances} & \multicolumn{1}{c}{} \\
\multicolumn{1}{c}{2} & \multicolumn{1}{c}{ $f_2(1270)$} & \multicolumn{1}{c}{5838.8/330=17.69} \\
\multicolumn{1}{c}{3} & \multicolumn{1}{c}{ $f_2(1430)$} & \multicolumn{1}{c}{327.39/330=0.9921} \\
\multicolumn{1}{c}{4} & \multicolumn{1}{c}{ $f_2(1525)$} & \multicolumn{1}{c}{336.97/330=1.0211} \\
\multicolumn{1}{c}{5} & \multicolumn{1}{c}{ $f_2(1640)$} & \multicolumn{1}{c}{327.76/330=0.9932} \\
\multicolumn{1}{c}{6} & \multicolumn{1}{c}{ $f_2(1810)$} & \multicolumn{1}{c}{348.07/330=1.0548} \\
\multicolumn{1}{c}{7} & \multicolumn{1}{c}{ $f_2(1910)$} & \multicolumn{1}{c}{327.36/330=0.9920} \\
\multicolumn{1}{c}{8} & \multicolumn{1}{c}{ $f_2(1950)$} & \multicolumn{1}{c}{327.26/330=0.9917} \\
\multicolumn{1}{c}{9} & \multicolumn{1}{c}{ $f_2(2010)$} & \multicolumn{1}{c}{326.68/330=0.9899} \\
\multicolumn{1}{c}{10} & \multicolumn{1}{c}{ $f_2(2150)$} & \multicolumn{1}{c}{368.81/330=1.1176} \\
\multicolumn{1}{c}{11} & \multicolumn{1}{c}{ $f_2(2300)$} & \multicolumn{1}{c}{326.54/330=0.9895} \\
\multicolumn{1}{c}{12} & \multicolumn{1}{c}{ $f_2(2340)$} & \multicolumn{1}{c}{326.64/330=0.9898} \\
\hline
\hline
\end{tabular}
\caption{Values of $\chi^2$ after fitting with omitted some specific resonance state 
in the $D0$ wave.}
\label{TableChi2-DN}
\end{table}

\subsection{Full \mbox{\boldmath $\pi\pi$} amplitude}

The set of all important partial-wave amplitudes modified and re-fitted in 
the previous sections allowed us to construct the full $\pi\pi$ amplitude and 
to calculate the total and differential $\pi\pi \to \pi\pi$ cross section up to about 2~GeV. 
In the following we briefly summarize the basic formulas to clarify the normalization 
and the partial-wave decomposition.

The partial-wave amplitudes are related to the phase shift and inelasticity as 
\begin{equation}
t_{\ell}^I(s) = \frac{\sqrt{s}}{4ik_1}\, {\rm [\eta_{\ell}^I(s)\;e^{2i\delta_{\ell}^I(s)} -1]}
\end{equation}
and summed with the Legendre polynomials $P_{\ell}(cos\theta)$ give the full 
invariant $\pi\pi$ amplitude ${\cal T}^I(s,t)$ in a given isospin channel $I$
\begin{equation}
{\cal T}^I(s,t) = 32\pi\sum_{\ell}\,(2{\ell}+1)\,t_{\ell}^I(s)\,P_{\ell}(cos\theta),
\label{partial}
\end{equation}
where $\theta$ is the 
scattering angle between two pions in the c.m. frame. 

 The full invariant amplitude for the $\pi^+\pi^-$ scattering is
\begin{equation}
{\cal T}_{\pi^+\pi^-}(s,t) = \frac{1}{3}\,{\cal T}^0(s,t) + \frac{1}{2}\,{\cal T}^1(s,t) + 
\frac{1}{6}\,{\cal T}^2(s,t) 
\label{pi-plus-pi-minus}
\end{equation}
which gives the differential cross section in the c.m. frame
\begin{equation}
\frac{d\sigma_{\pi^+\pi^-}}{d\Omega} = \frac{1}{64\pi^2s}|{\cal T}_{\pi^+\pi^-}|^2\,,
\label{differential}
\end{equation}
and the total cross section 
\begin{equation}
\sigma^{tot}_{\pi^+\pi^-}(s) = \frac{{\rm Im} {\cal T}_{\pi^+\pi^-}(s,\theta\!=\!0)}{2k_1\sqrt{s}}\,.
\label{pi-plus-pi-minus}
\end{equation}

\begin{figure}[h!]
\includegraphics[angle=270,scale=0.47]{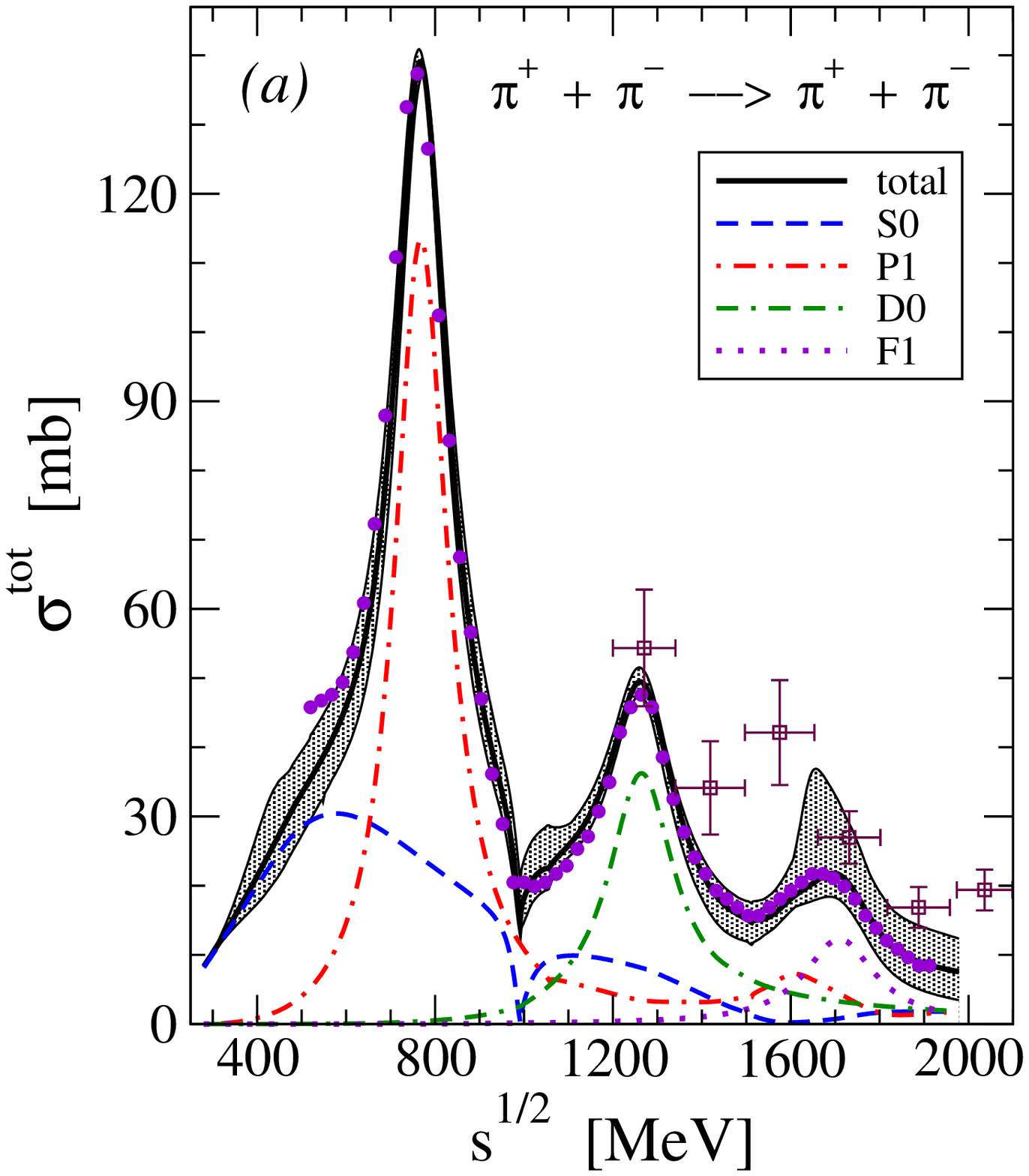}
\includegraphics[angle=270,scale=0.47]{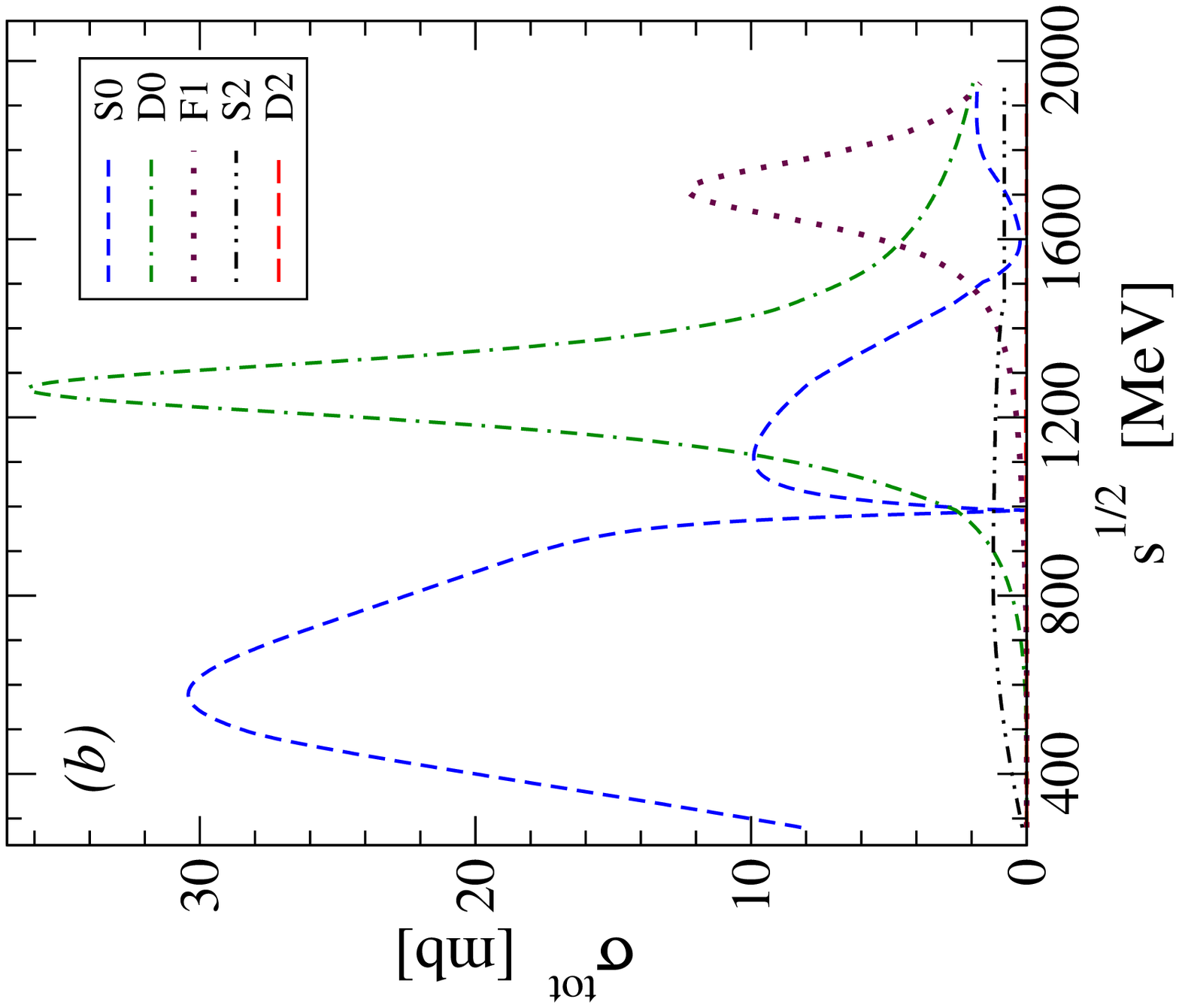}
\caption{Prediction of the total cross section in the low-energy 
region is shown for the $\pi^+\pi^- \to \pi^+\pi^-$ scattering. Presented are 
the contributions from most important partial-waves ({\it a}) and also for 
weaker amplitudes ({\it b}). The gray band on ({\it a}) represents uncertainties 
of the total cross section. Data are from \cite{Hyams} (dots) 
and \cite{Biswas} (the six points above 1200~MeV with bigger errors).}
\label{fig-total}
\end{figure}

The total cross section and its components from the individual partial-waves are presented 
on Fig. \ref{fig-total}. 
%
%
Well seen is the dominance of the $f_0(500)$ and $\rho(770)$ below 1~GeV. 
Two maxima above this energy are formed mostly by $f_2(1270)$ and $\rho_3(1690)$.
Other waves - $S2$ and $D2$ have minor influence on the cross section.
Our theoretical predictions very well agree with data from \cite{Hyams,Biswas}.
Worthy is to notice that such good agreement has been achieved without fitting directly to these data. 
Uncertainties of the total cross sections presented on the figure were calculated using 
the Monte Carlo method for 1000 randomly generated sets of all free, in minimizations, 
parameters taken within their 1~$\sigma$ deviation.

\begin{figure}[h!]
\includegraphics[angle=270,scale=0.45]{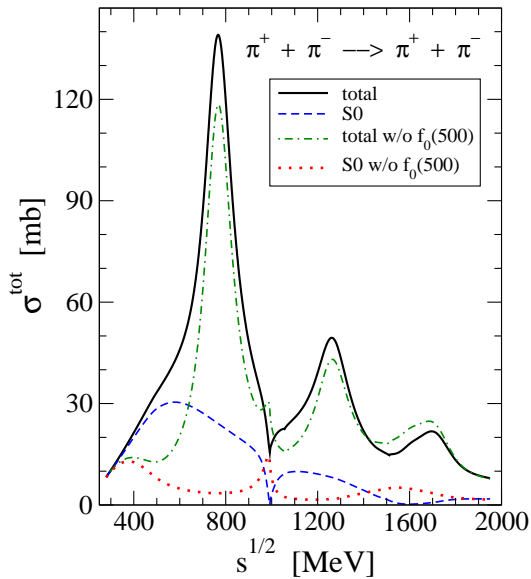}
\caption{Prediction of the total and $S$-wave cross sections in the low-energy 
region with and without $f_0(500)$ are shown for the $\pi^+\pi^- \to \pi^+\pi^-$ scattering.}
\label{fig-total-nosigma}
\end{figure}

In many experimental analyses performed in the seventies (see some papers in \cite{Hyams,ExpData}) 
the $f_0(500)$ was replaced by a very broad and heavier (with mass around 1~GeV) resonance.
Its influence on the phase shifts and cross sections was weaker and distributed in a wider energy range.
Therefore, it is interesting to see how the cross section would look if the $f_0(500)$ has 
been removed from the analysis.
We demonstrate it on Fig. \ref{fig-total-nosigma}. 
Of course the main difference, in comparison with Fig. \ref{fig-total}, concerns region below 0.7~GeV 
which is almost completely determined by the $f_0(500)$. 
Small enhancement near the \pipi threshold is caused by the fixed scalar-isoscalar scattering 
length and the slope parameter. It is worth to pay attention on completely different behavior of 
the cross section on Figs. \ref{fig-total} and \ref{fig-total-nosigma} near the \KK threshold 
where one observes the $f_0(980)$. 
In the cross sections without the $f_0(500)$ clearly seen is a small peak instead of the 
deep minimum seen on Fig. \ref{fig-total}. 
The peak appears due to absence of about 90 degree component in the $S$ wave phase shift generated by the $f_0(500)$ pole and zero.
On Fig. \ref{fig-total-nosigma} one can also see that the interference of the $f_0(500)$ with all other amplitudes is positive 
in the whole energy range except of the region between around 1.5~GeV and 1.7~GeV and vicinity of the small peak around 1~GeV.

\begin{figure}[h!]
\includegraphics[angle=270,scale=0.33]{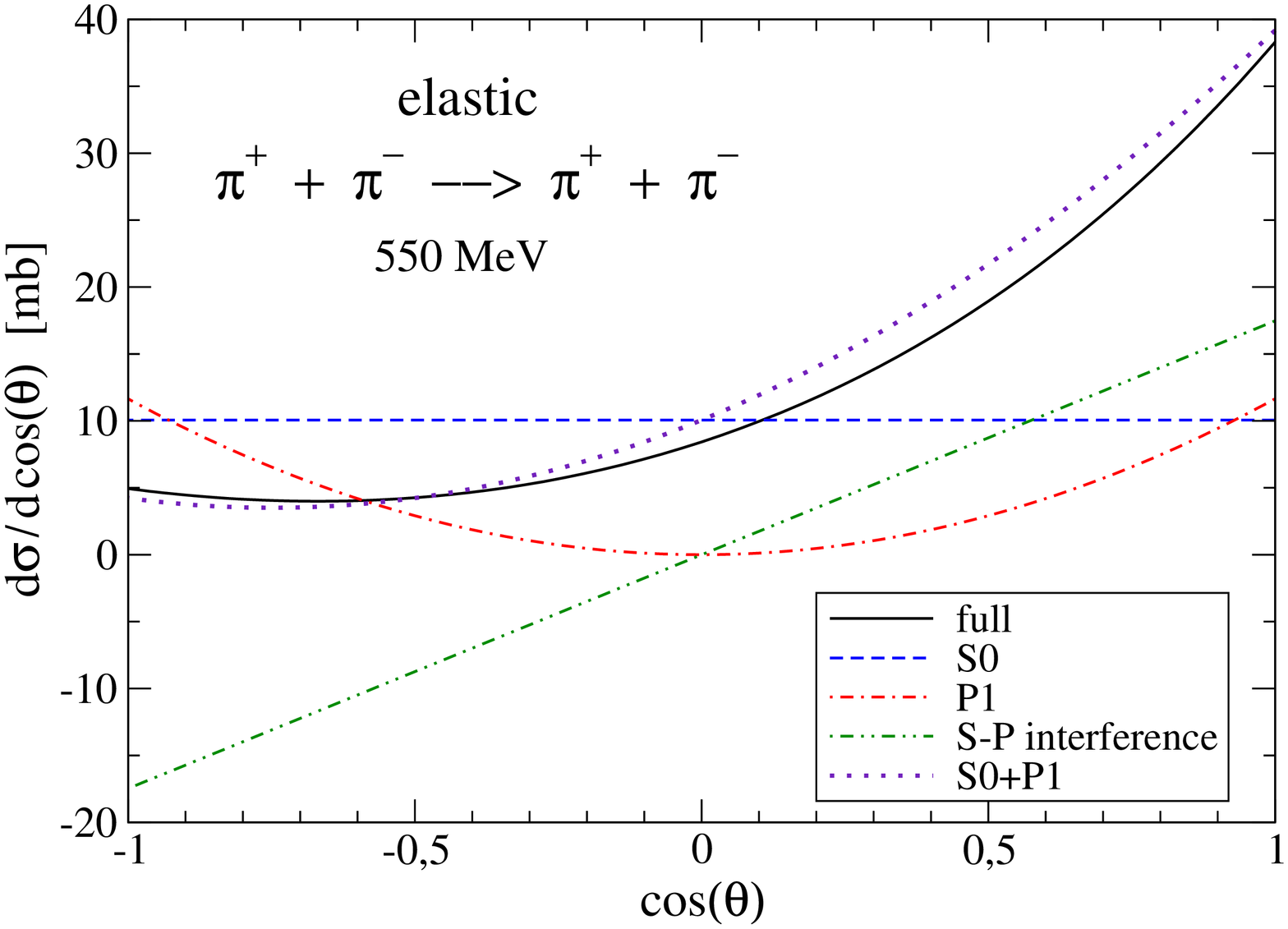}
\includegraphics[angle=270,scale=0.33]{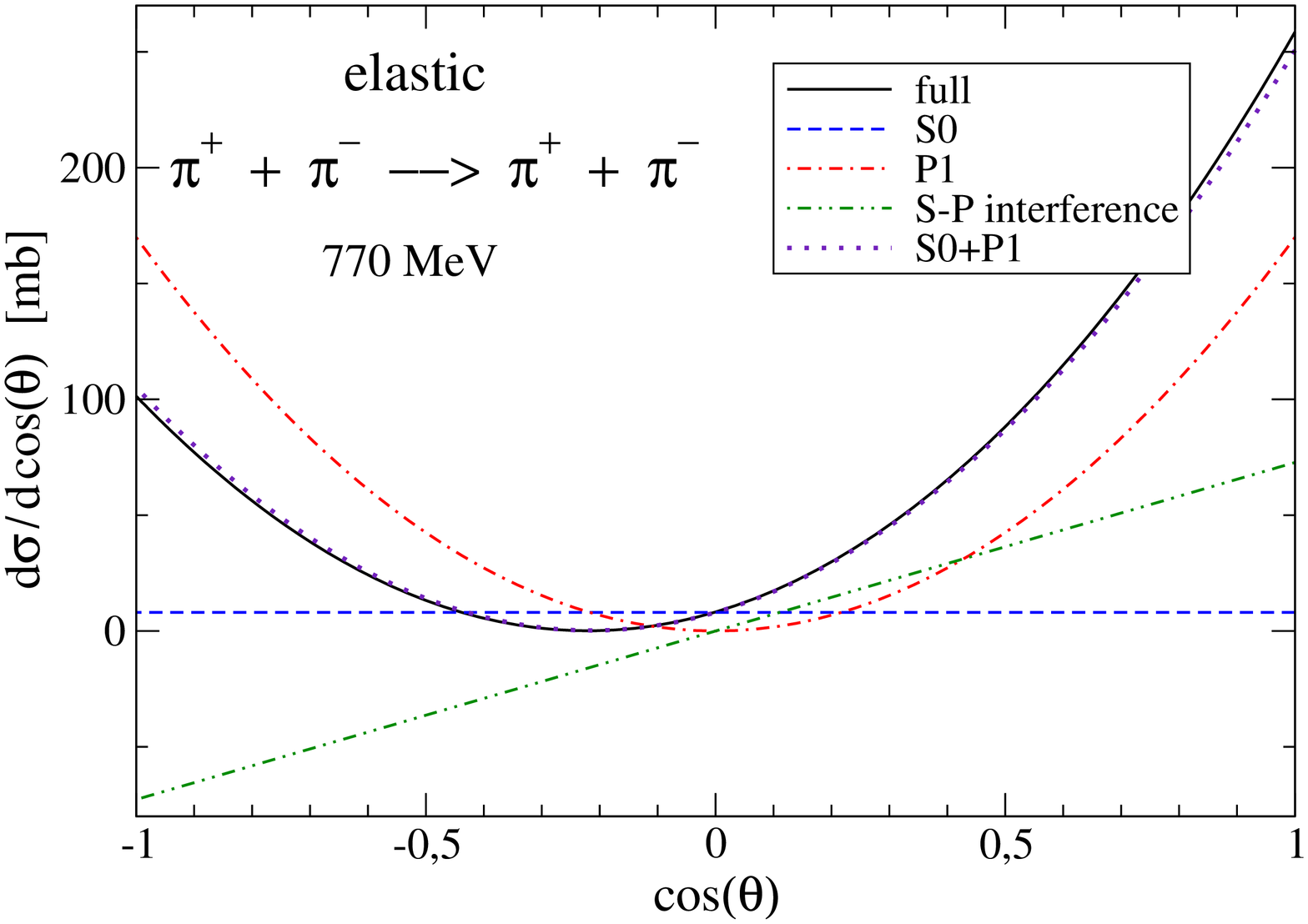}
\caption{Angular dependence of the cross section for two energies is shown 
as predicted in the $\pi^+\pi^-$ scattering. Contributions from 
particular partial-waves are shown.}
\label{fig-differential}
\end{figure}

The elastic differential cross sections at two energies: 550~MeV and 770~MeV are presented on Fig. \ref{fig-differential}.
Clearly seen is, as expected, the significant role of the $S$-wave at 550~MeV (dominated by the $f_0(500)$). 
An interference between the $S$ and $P$ waves  producing the enhancement at forward angles
is substantial and interference with other waves - only  noticeable.
In vicinity of the $\rho(770)$ the cross section is, of course, dominated by the $P$-wave. 
Again, the interference between the $S$ and $P$ waves is very important and interference with other waves barely noticeable.

\section{Conclusions}

We have constructed new multi-channel $D0$- and $F1$-wave $\pi\pi$ amplitudes 
using the multi-channel Breit-Wigner formalism. 
When constructing the $D0$ wave we utilized the tensor-isoscalar resonances 
presented in the latest issue of the PDG tables.
We showed that the data in the $F1$-wave can be satisfactorily described 
considering only one resonance $\rho_3(1690)$. The other state $\rho_3(1990)$ 
also listed in PDG appeared as unnecessary in the data description. 

The previously constructed $S0$- and $P1$-wave multichannel amplitudes 
and the new ones, $D0$ and $F1$, were modified in the dispersive analysis 
using the Roy-like equations (GKPY). The isotensor amplitudes 
$S2$ and $D2$, also used in the analysis, were taken in the phenomenological 
form and were not changed.

The modified (re-fitted) partial-wave multi-channel amplitudes $S0$, $P1$, $D0$, and $F1$ 
are optimized to the experimental data in the considered channels and they fulfill 
the crossing symmetry condition imposed by the GKPY dispersion equations. 
The overall description of the data is satisfactory. In the $S0$-wave amplitude 
a position of the $f_0(500)$ pole on the second Riemann sheet was changed to the value 
very well consistent with the PDG values and with our previous result.
In the analysis of the $D0$ wave we concluded that apart of the dominant $f_2(1270)$ state 
the resonances $f_2(1525)$, $f_2(1810)$, and $f_2(2150)$ also do play an important role 
in the data description and are also required in the dispersive analysis.

The modified partial-wave amplitudes were utilized in constructing the full invariant 
$\pi\pi$ scattering amplitude and the cross sections were calculated.

\section*{ACKNOWLEDGEMENT}
We want to thank Yurii S. Surovtsev for his help and many useful 
discussions with him. We are particularly grateful to George Rupp for many 
important comments and observations, and for very fruitful discussions with 
one of us.
This work has been partially supported by the Polish Science
Center (NCN) Grants No. Dec-2013/09/B/ST2/04382 and DEC-2014/15/N/ST2/03504 
and by the Grant Agency of the Czech Republic under the grant No. P203/15/04301.


\end{document}